\pdfoutput=1
\documentclass[iop,apj]{emulateapj}

\def\lessim{\mathrel{\hbox{\rlap{\hbox{\lower4pt\hbox{$\sim$}}}\hbox{$<$}}}}
\def\grtsim{\mathrel{\hbox{\rlap{\hbox{\lower4pt\hbox{$\sim$}}}\hbox{$>$}}}}

\begin{document}

\title{Exploring the Role of Globular Cluster Specific Frequency on the Nova Rates in Three Virgo Elliptical Galaxies}
\author{C. Curtin\altaffilmark{1,2}, A. W. Shafter\altaffilmark{1}, C. J. Pritchet\altaffilmark{3}, J. D. Neill\altaffilmark{4}, A. Kundu\altaffilmark{5,6}, T. J. Maccarone\altaffilmark{7}}
\altaffiltext{1}{Department of Astronomy and Mount Laguna Observatory, San Diego State University, San Diego, CA 92182}
\altaffiltext{2}{Centre for Astrophysics and Supercomputing, Swinburne University of Technology, Hawthorn, VIC 3122, Australia}
\altaffiltext{3}{Department of Physics and Astronomy, University of Victoria, Victoria, BC V8W 2Y2}
\altaffiltext{4}{Space Radiation Laboratory 290-17, California Institute of Technology, Pasadena, CA 91125, USA}
\altaffiltext{5}{Tata Institute of Fundamental Research, Mumbai 400005, India}
\altaffiltext{6}{Eureka Scientific Inc., 2452 Delmer St, Suite 100, Oakland CA 94602, USA}
\altaffiltext{7}{Department of Physics, Texas Tech University, Box 41051, Lubbock TX 79409-1051, USA}

\begin{abstract}

It has been proposed that a galaxy's nova rate might be enhanced by the
production of nova progenitor binaries in the dense cores of its
globular clusters (GCs).
To explore this idea, relative nova rates in
three Virgo elliptical
galaxies, M87, M49 and M84, which have
significantly different GC specific frequencies
($S_{N}$) of 14, 3.6, and 1.6, respectively,
were measured over the course of
4 epochs spanning a period of 14 months. To simplify
the analysis, observations of the nearly equidistant galaxies
were made on the same nights, with the same integration times, and
through the same filter (H$\alpha$),
so that the relative numbers of novae discovered
would reflect the relative nova rates.
At the conclusion of our survey
we found a total of 27 novae associated with M87, 37 with M49, and 19 with M84.
After correcting for survey
completeness, we found
annual nova rates of $154^{+23}_{-19}$, $189^{+26}_{-22}$,
and $95^{+15}_{-14}$, for M87, M49, and M84, respectively, corresponding to
$K$-band luminosity-specific nova rates of $3.8\pm1.0$, $3.4\pm0.6$,
and $3.0\pm0.6$
novae per year per $10^{10}~L_{K,\odot}$.
The overall results of our study suggest that a galaxy's nova
rate simply scales with its luminosity, and is insensitive
to its GC specific frequency.
Two novae, one in M87 and one in M84, were found
to be spatially coincident with known GCs. After correcting for the
mass fraction in GCs, we estimate that novae are likely enhanced
relative to the field by at least
an order of magnitude in the GC systems of luminous
Virgo ellipticals.

\end{abstract}


\section{Introduction}

Classical novae are a subclass of cataclysmic variables in which
a progenitor system increases in brightness by $\sim$10--20 mag
over a period of days to weeks and then returns to its
pre-outburst magnitude over a period of days to years.
The progenitor system is a short period,
semi-detached binary consisting of a white dwarf
primary accreting material from a Roche lobe filling late-type companion.
If the accretion rate is sufficiently low,
the accreted material forms a thin shell of degenerate matter on the surface of the white dwarf. Once the temperature and pressure at the base of the accreted envelope
become sufficiently high, a thermonuclear runaway ensues,
resulting in a nova eruption. Novae can reach an absolute magnitude,
$M_{V}\approx-10$, making them among the most luminous explosions
in the Universe, easily visible in galaxies
as distant as the Virgo cluster \cite[see][for a review]{sha08}.  

The nature of the progenitor system suggests that novae will be
recurring events. Once a nova eruption occurs the material will begin to
build up again and the process will repeat. Recurrence times
can be as short as 1 year \citep[e.g., see][]{dar14,hen14} all the way up to $10^5$ yr, or
perhaps longer. Despite the fact that all
novae are believed to be recurrent, only
novae with more than one recorded outburst
are referred to as ``recurrent novae".
Over time, the orbital period of a nova progenitor
binary decreases as a result of
angular momentum losses due to magnetic
stellar winds from the tidally-locked donor star, or from
gravitational wave radiation, or both. As the orbital period decreases,
so does the average
accretion rate onto the white dwarf primary, resulting in an increase
in the recurrence time \citep{ibe92}.
In addition,
in a given stellar population, the average mass of the primaries in nova
systems are expected to
decrease with increasing time elapsed since the zero-age main sequence
population formed \citep{tut95}.
A less massive primary needs to accrete more material to reach the
necessary temperature near its surface to undergo a thermonuclear runaway
\citep[e.g., see][]{tow05}.
Such systems will undergo less frequent (and dimmer) nova outbursts compared
to systems with higher mass primaries \citep{rit91,liv92,kol95}.

Accounting for the above factors,
population synthesis models have predicted that $K$-band
luminosity-specific nova rates, $\nu_K$,
should be significantly higher in the younger populations found in spiral
disks compared with older populations found in elliptical galaxies and
in the bulges of spiral galaxies \citep[e.g.,][]{yun97}.
Despite these predictions,
many nova surveys have found that, within observational uncertainties
(which can be large), $\nu_K$ appears to be relatively
independent of Hubble type
\citep[e.g.][]{cia90a, fer03, wil04, coe08, gut10, fra12}, although
some surveys have
found evidence that $\nu_K$ varies between galaxies,
though not necessarily as predicted by models
\citep[e.g.][]{del94, sha02, mad07}.
In addition,
nova surveys of M31, in which nearly 1000 novae have been observed
\citep[e.g., see][]{pie07}, have repeatedly found that the nova
spatial distribution is more centrally concentrated than the
background light, indicating a higher $\nu_K$ in the bulge than
in the disk \citep[e.g.][]{cia87, cap89, sha01, dar04, dar06}.
More recently \citet{nei04} have found a similar result for M81.
In light of the apparent discrepancy between theory and
observation found in their early M31 survey \cite{cia87}
suggested that a significant number of nova progenitor
binaries may have formed
in M31's globular cluster (GC) population and been subsequently
injected into the bulge through 3-body encounters within clusters, or
through tidal disruptions of entire clusters, or both,
thereby enhancing the bulge nova rate.

It has long been known that the number of Galactic low-mass X-ray binaries
per unit mass are enhanced in GC environments \citep{cla75,kat75}, and
it has been suggested that many Galactic bulge X-ray sources could
have had their origin in GCs \citep{gri88}.
A similar enhancement of X-ray
sources has also been seen in M31's GC population \citep{cra84, dis02}.
In recent years, it has been suggested that a significant population of
close binaries with white dwarf accretors (e.g., nova progenitors)
could also be formed in the dense
cores of GCs \citep[e.g.,][]{poo06}.
Observations by the {\it Hubble Space Telescope\/} ({\it HST\/}) and
Chandra had already begun to reveal
these binaries in Galactic GCs
\citep[e.g.,][]{edm03,hei03,kni02,poo02}, although there is no
compelling evidence that they are enhanced relative to the field.

Additional evidence for the idea that nova progenitors could be
produced in GCs has come from the study of nova rates
in Virgo elliptical galaxies. \cite{sha02} reported that
the nova rate in the giant elliptical galaxy
M87 could be as high as $\sim300$~yr$^{-1}$, while
for the slightly more luminous elliptical, M49,
\cite{fer03} found a rate of $R\approx100$. It is interesting that,
like the ratio of observed nova rates,
M87 has a specific frequency of GCs about three times that
of M49 \citep{bro06}. Although, these results are consistent with the
possibility that the large GC population of M87 enhances its nova rate,
absolute nova rates extrapolated from observed nova rates
are largely uncertain given the small spatial coverage
of these HST surveys, and comparisons of these rates between
surveys can be unreliable.

In 2011 we began a definitive test of the putative
relationship between a galaxy's GC specific frequency, $S_{N}$,
and its luminosity-specific nova rate, $\nu_K$.
Specifically, we searched for novae in three essentially equidistant
Virgo elliptical galaxies having different GC specific
frequencies. The galaxies were observed
under nearly identical conditions using the same
telescope, observing cadence, filters, and exposure times.
With this approach we were able to simply compare the
relative numbers of novae discovered in each
of the three galaxies directly to estimate relative nova rates, while
avoiding the large
uncertainties that result when attempting
to measure absolute nova rates for different galaxies in different surveys.
Preliminary results of our program
were discussed in \cite{sha13} and \cite{me}. Here, we present
the final results of the first phase of our ongoing Virgo galaxy nova study.

\section{Observations}

We chose three Virgo ellipticals for our study:
M87 (NGC 4486), M49 (NGC 4472) and M84 (NGC 4374). As members of the Virgo
cluster, these three galaxies are approximately equidistant. They also have
similar luminosities,
but they have widely varying GC specific frequencies, $S_{N}$
(see Table~\ref{tab1}).
We observed all three galaxies over four epochs from February 2011 to March 2012
using the Canada-France-Hawaii 3.6-m telescope (CFHT).
All observations were made using the MegaCam which consists of a
mosaic of 36 2048$\times$4612 pixel CCDs covering a full field of
view of $\sim1$~deg$^2$ at a resolution of 0.187$''$/pixel \citep{bou03}.
Exposure times were kept nearly constant between galaxies,
varying only slightly due to weather in a small number of cases
to ensure that the derived relative nova rates would be directly comparable.
The observations are summarized in Table~\ref{tab2}. 

\begin{deluxetable}{lccccc}
\tablenum{1}
\label{tab1}
\tablecaption{Galaxy Properties}
\tablewidth{0pt}
\tablecolumns{6}
\tablehead{
\colhead{} & \colhead{$S_N$} & \colhead{$(m-M)$} & \colhead{$B_{T,0}$} & \colhead{$(B-V)_0$} & \colhead{$(V-K)_0$}\\  \colhead{Galaxy} & \colhead{} & \colhead{(mag)} & \colhead{(mag)} & \colhead{(mag)} & \colhead{(mag)}
}

\startdata

M87 &             14 &           31.03&  9.49&           0.96&           3.21    \\
M49 &             3.6&           31.06&  9.28&           0.96&           3.34    \\
M84 &             1.6&           31.32&  9.99&           0.87&           3.28    \\

\enddata

\tablecomments{
$S_N$ are from \cite{bro06},
$(m-M)$ are from \cite{ton01},
$B_{T,0}$ and $(B-V)_0$ are from \cite{dev91}, and
$(V-K)_0$ are from \cite{fro78}.}

\end{deluxetable}

\begin{deluxetable}{lccccc}
\tablenum{2}
\label{tab2}
\tablecaption{Observations}
\tablewidth{0pt}
\tablecolumns{6}
\tablehead{
\colhead{Epoch} & \colhead{Date} & \colhead{Galaxy} & \colhead{Exp} & \colhead{Seeing} & \colhead{Weather}
}

\startdata

1    &  2011 Feb 02  & M49 & $12\times5$m& 0.99$''$ & phot\\
\dots&  2011 Feb 02  & M84 & $12\times5$m& 0.82$''$ & phot\\
\dots&  2011 Jan 30  & M87 & $12\times5$m& 0.87$''$ & phot\\
2    &  2011 Mar 30  & M49 & $12\times5$m& 0.82$''$ & phot\\
\dots&  2011 Apr 02  & M84 & $12\times5$m& 0.89$''$ & phot\\
\dots&  2011 Mar 29  & M87 & $16\times5$m& 0.95$''$ & p cldy\\
3    &  2012 Feb 26  & M49 & $11\times5$m& 1.19$''$ & phot\\
\dots&  2012 Feb 27  & M84 & $12\times5$m& 0.72$''$ & p cldy\\
\dots&  2012 Feb 26  & M87 & $12\times5$m& 0.96$''$ & phot\\
4    &  2012 Mar 21  & M49 & $15\times5$m& 0.72$''$ & p cldy\\
\dots&  2012 Mar 22  & M84 & $12\times5$m& 0.66$''$ & p cldy\\
\dots&  2012 Mar 21  & M87 & $13\times5$m& 0.81$''$ & p cldy\\

\enddata

\end{deluxetable}

We chose to image
all three galaxies through a narrow-band H$\alpha$ filter redshifted to the
mean velocity of the Virgo cluster
($\lambda_c=6584$\AA, FWHM = 76\AA [$\sim3500$~km~s$^{-1}$]). For nova surveys
imaging in H$\alpha$ has two principal
advantages over broad-band imaging.
Novae exhibit strong H$\alpha$ emission lines shortly after outburst
that fade slowly relative to the continuum,
typically requiring months to decline by more than 2 mag.
The slow decay rate in H$\alpha$ means that we can detect novae over a longer
time interval, requiring
less frequent temporal sampling. Another advantage of H$\alpha$
over broad-band observations, is that
H$\alpha$ images provide a greater contrast between novae and the bright
galaxy background, resulting in increased survey completeness,
particularly near the bright nuclei of galaxies \citep[e.g.,][]{cia87}.

Based on the luminosity function of novae in both M31 and the Galaxy,
we estimate that approximately 50\% of the Virgo novae should reach a
peak absolute magnitude, $M_{max,{\rm H}\alpha}=-7.5$\footnote{
The H$\alpha$ magnitude is defined on the AB system where
$m_{{\rm H}\alpha} = 0$ for $f_{\lambda} = 2.53 
\times 10^{-9}$~ergs~cm$^{-2}$~s$^{-1}$~\AA$^{-1}$.
Through a 76\AA\
wide filter, this corresponds to a zero point
flux of $\sim1.9\times10^{-7}$~ergs~cm$^{-2}$~s$^{-1}$.}.
Assuming a distance modulus to the Virgo cluster of
$\mu_0 \simeq 31.0$ (see Table~\ref{tab1}),
we require observations that can reach
$m_{{\rm H}\alpha} \simeq 23.5$.
Numerical simulations revealed that with
a mean seeing of 0.8$''$ typical for the CFHT we could detect
a $m=23.5$ nova during grey time
with a signal-to-noise ratio, $S/N$, $\sim12$ at $60''$ from the center of 
its host galaxy and $S/N\sim6$ at $30''$ from center in a one-hour exposure.
To avoid saturation of the nucleus, and
to enable the removal of cosmic rays hits, our 1~hr exposures
were divided into 12 five-minute integrations.

Our complete survey was spread over four epochs of observations spanning
two years.
The first two epochs were separated by two months, which, given the persistence
of the nova H$\alpha$ emission, provides reasonable temporal coverage, with
minimal overlap of detected novae.
The third and fourth epochs were scheduled in the following year
to ensure that we could discriminate between slowly evolving novae and other
types of variable sources.

Given an estimate for the absolute nova rate in M49
\citep[100~yr$^{-1}$,][]{fer03}, we conducted a simulation to estimate
the number of novae we could expect to observe in each epoch.
In the simulation, we consider \citet{dev48} model for the
brightness profile of the galaxy and then distribute
artificial novae with a density that follows the background light.
The peak brightnesses and fade rates for the artificial novae are based on
H$\alpha$ light curves of 14 novae observed in
M31 \citep{sha01} and M81 \citep{nei04},
adjusted to the distance of the Virgo cluster.
Given the temporal sampling,
exposure times (see Table~\ref{tab2}) and an estimate of the nova
rate
we compute the number of novae expected to be observed above a given $S/N$.
Under the null hypothesis that the nova rate is independent of GC
specific frequency, and scales directly with K-band luminosity,
we would expect to observe 22, 32 and 15 novae in M87, M49, and M84,
respectively with a $S/N \ge 10$.
On the other hand,
in the extreme (and likely unrealistic) case that {\it all\/} nova progenitors
are formed in GCs, the number of novae expected scales directly
with the GC specific frequency and
the predicted rates were 86, 32 and 7 novae
for the three galaxies, respectively.

\section{Data Reduction and Analysis}

In each observation, we positioned the target galaxy on the top
of CCD chip 22 near the center of the MegaCam array.
More than 99\% of the total light from each galaxy falls on the
same six central chips surrounding the nucleus
(chips 12, 13, 14, 21, 22, 23),
and these chips were the focus of our analysis.
The images were not significantly dithered,
and so chip gaps of $\sim13''$ persist in the final images for each epoch.
The area obscured by chip gaps is $\sim3\%$ of the total survey area.
However, due to the positioning of the target galaxies,
no chip gaps are encountered within a radius of $120''$ of the nuclei.
Using the same photometry used to plot the cumulative background
light in Figure 2, we estimate the light enclosed within this radius
to be greater than 60\% of the total light for all targets,
leaving considerably less than $1.2\%$ of the light (and likely the novae)
obscured by chip gaps. This loss is negligible compared to other
uncertainties inherent in nova rate calculations,
and we have not made any formal correction for it in the analysis.

The MegaCam data come pre-processed with bad pixel masking,
bias subtraction, and flat fielding being performed prior to the data
release. Once the pre-processed data were in hand, individual 5 minute
exposures for each of the six central chips were spatially registered using the
IRAF\footnote{IRAF (Image Reduction and Analysis Facility)
is distributed by the National Optical Astronomy Observatory,
which is operated by Association of Universities for
Research in Astronomy, Inc., under cooperative agreement with the
National Science Foundation} routine {\tt WREGISTER}.
The individual CCD images were then median stacked using the tasks
{\tt IMALIGN\/} and {\tt IMCOMBINE\/} to produce master images for each chip
in each epoch, typically representing 60~min of total exposure time
(see Table~\ref{tab2}).
Finally, the WCS for each master image was recalibrated to match
between epochs using the IRAF task {\tt CCMAP\/} in conjunction with
precise coordinates
for several reference stars on each image obtained from the
US Naval Observatory (USNO) B1 catalog \citep{mon03}.  

\subsection{Nova Detection}

To identify nova candidates the master images were spatially aligned,
point-spread function (PSF) matched and differenced using the
ISIS package \citep{ala98}.
In the resulting images objects of variable brightness are conspicuous,
allowing us to detect them by eye.
Variables qualified as nova candidates if they had an approximately
Gaussian PSF, and if they were visible in only
one or two consecutive epochs, but did not appear on
both sides of the 10 month
gap in observations.

Within the inner $\sim30''$ of each galaxy center the differencing
became less effective due to the increasingly bright background light.
In an attempt to detect nova candidates in the inner regions we applied
the IRAF task {\tt MEDIAN\/}, which produced a smoothed approximation of the
background light profile.
Stellar sources are not completely removed in the
median smoothing process, but they are greatly reduced
in amplitude.
The resulting median image was then subtracted
from the original to produce a background-subtracted image of the
central region where stellar sources are easier to identify.
The background-subtracted images from separate epochs were then
compared to identify nova candidates.
No new nova candidates were identified using this method,
however we were able to confirm pre-existing candidates or
in a couple of cases eliminate candidates due to non-stellar PSFs
first identified in the background subtracted images.
All told, we have discovered a total of 27, 37, and 19 novae in the
fields of M87, M49, and M84.

\subsection{Photometry}

To calibrate the nova H$\alpha$ magnitudes, we began by performing
aperture photometry using the IRAF task {\tt PHOT\/}
to determine instrumental magnitudes for the nova candidates and
several nearby reference stars from the USNO B1 catalog, which
were employed as secondary standards.
For nova candidates close to the nucleus we needed to account for
the steepness of the background light profile before we could perform accurate
aperture photometry. To ensure that we were able to accurately measure the
nova brightness, we pre-processed subimages centered on the nova
using the task {\tt IMSURFIT\/}, which excludes a small circular region
surrounding the nova and then calculates and subtracts a fit of the
background light.

Once instrumental magnitudes were in hand,
standard AB magnitudes for the novae
(which correspond to monochromatic fluxes averaged over the
H$\alpha$ filter bandpass) were estimated
through differential photometry with respect to the
known $R$-band magnitudes of the reference stars.
Conversion to a true H$\alpha$ magnitude that would reflect
the total nova emission at H$\alpha$ would require a knowledge of the
width of the nova H$\alpha$ emission relative to the filter bandpass.
Since the former value varies from nova to nova depending on the
expansion velocity of the nova shell, an accurate measure
of a nova's H$\alpha$ flux through filter photometry
is not possible. For consistency with our
earlier studies \citep[e.g., see][] {cia90b,sha00},
we assume a 100\% filling fraction in the present analysis.
In this case, the adopted
nova H$\alpha$ magnitudes are simply equivalent to standard AB
magnitudes, with $m_{{\rm H}\alpha} = 0$ corresponding to
$f_{\lambda} = 2.53\times10^{-9}$~ergs~cm$^{-2}$~s$^{-1}$~\AA$^{-1}$.

\begin{figure}

\includegraphics[scale=0.35]{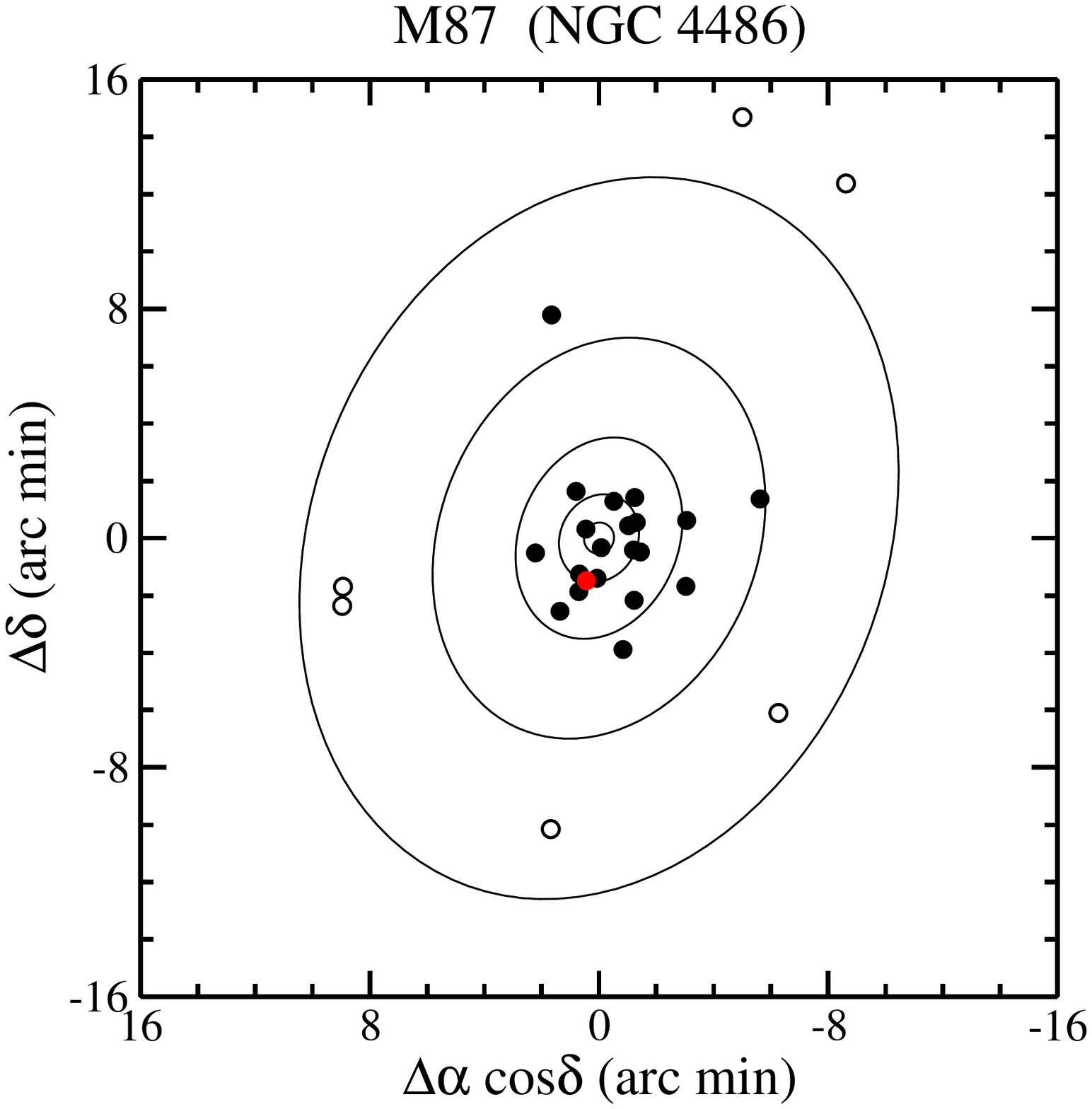} 
\includegraphics[scale=0.35]{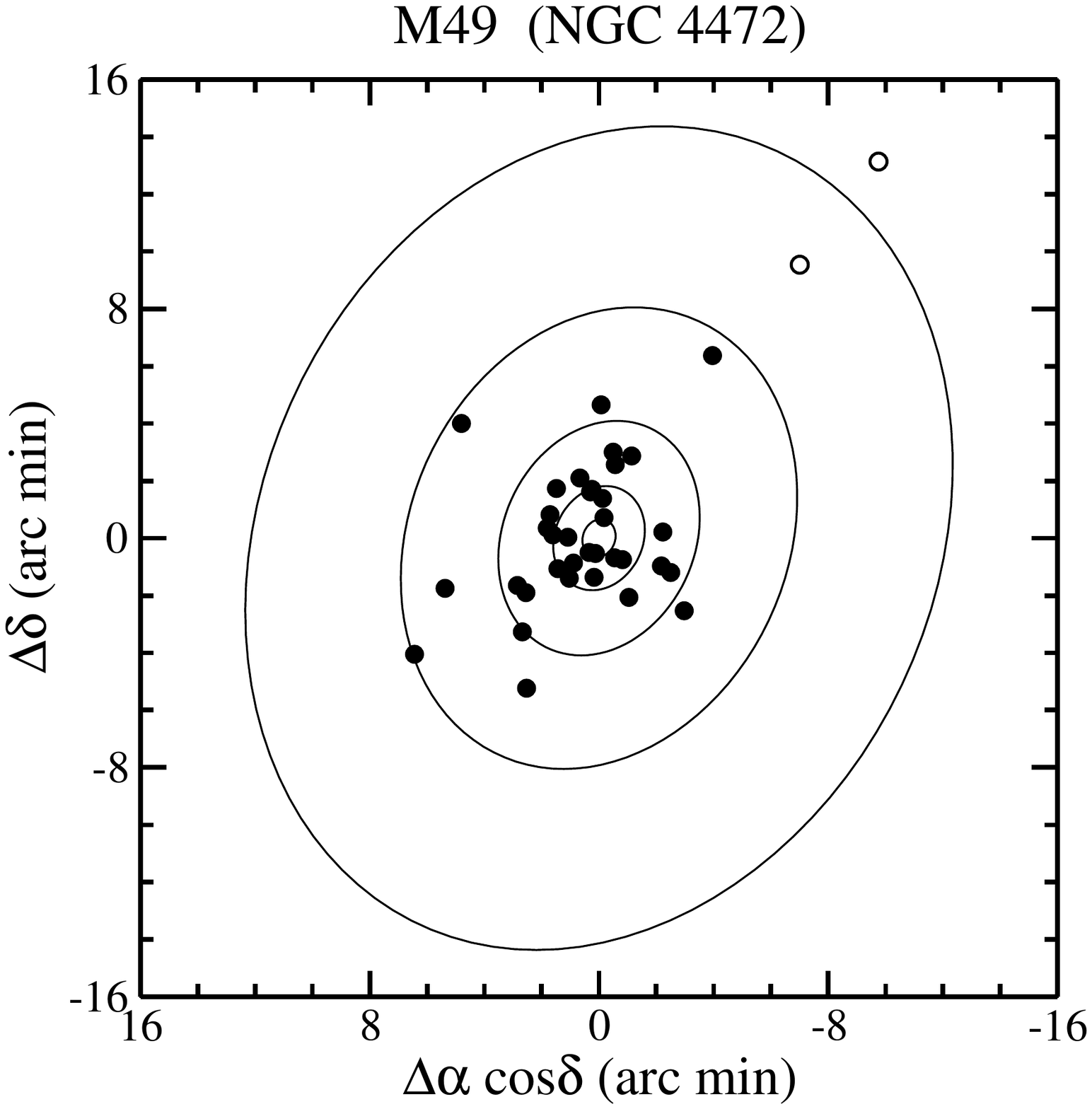} 
\includegraphics[scale=0.35]{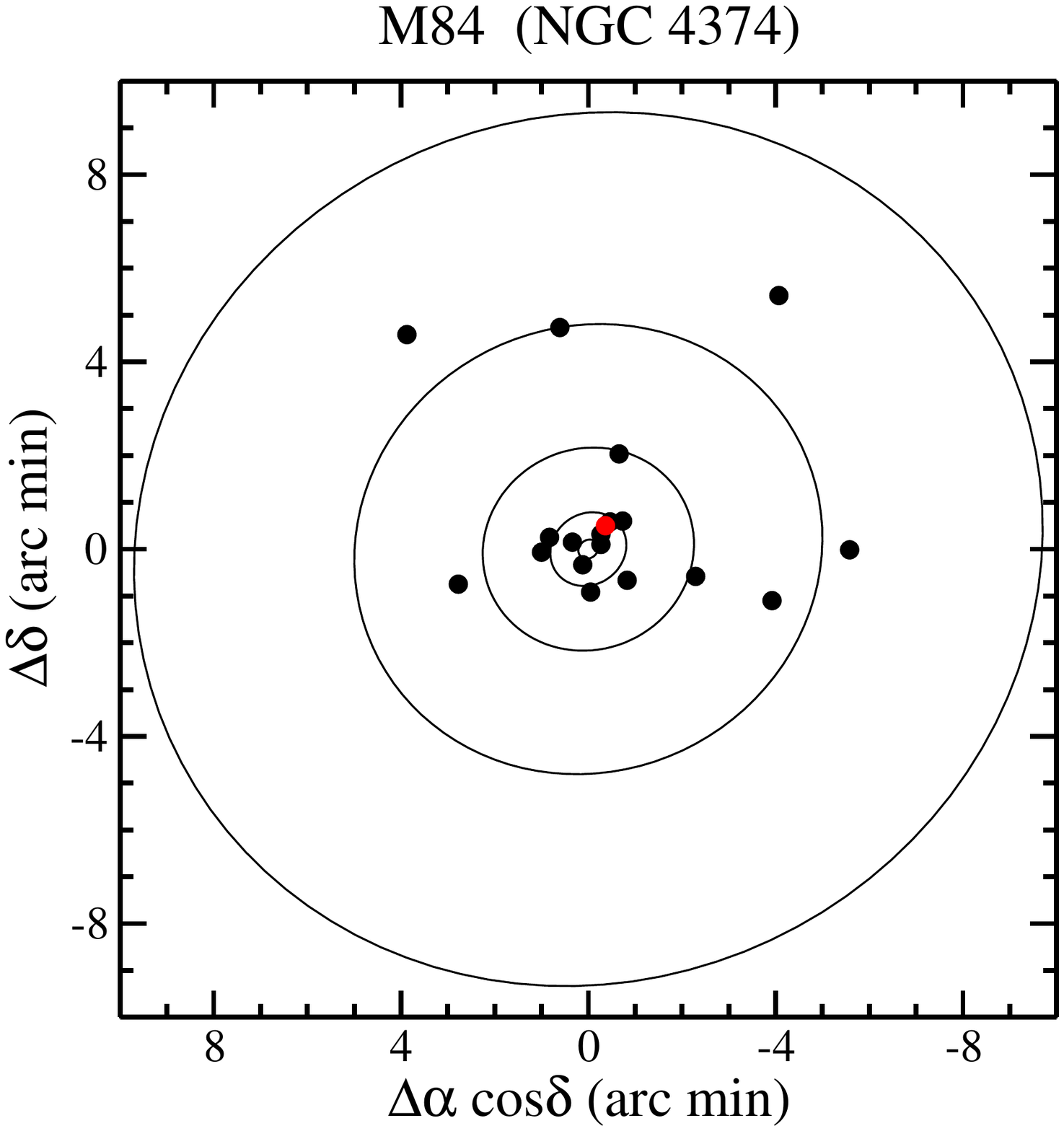} 

\caption{The spatial positions of the novae detected in each of our surveyed galaxies, plotted over equally spaced isophotes from $\mu=20$ to $\mu=28$ mag~arcsec$^{-2}$.
Open circles represent novae that are more than $10'$ from the nucleus, which have been excluded from the second analysis of M87.  Filled red circles mark the locations of GC novae.
\label{fig1}}

\end{figure}

\begin{figure}

\includegraphics[width=0.5\textwidth]{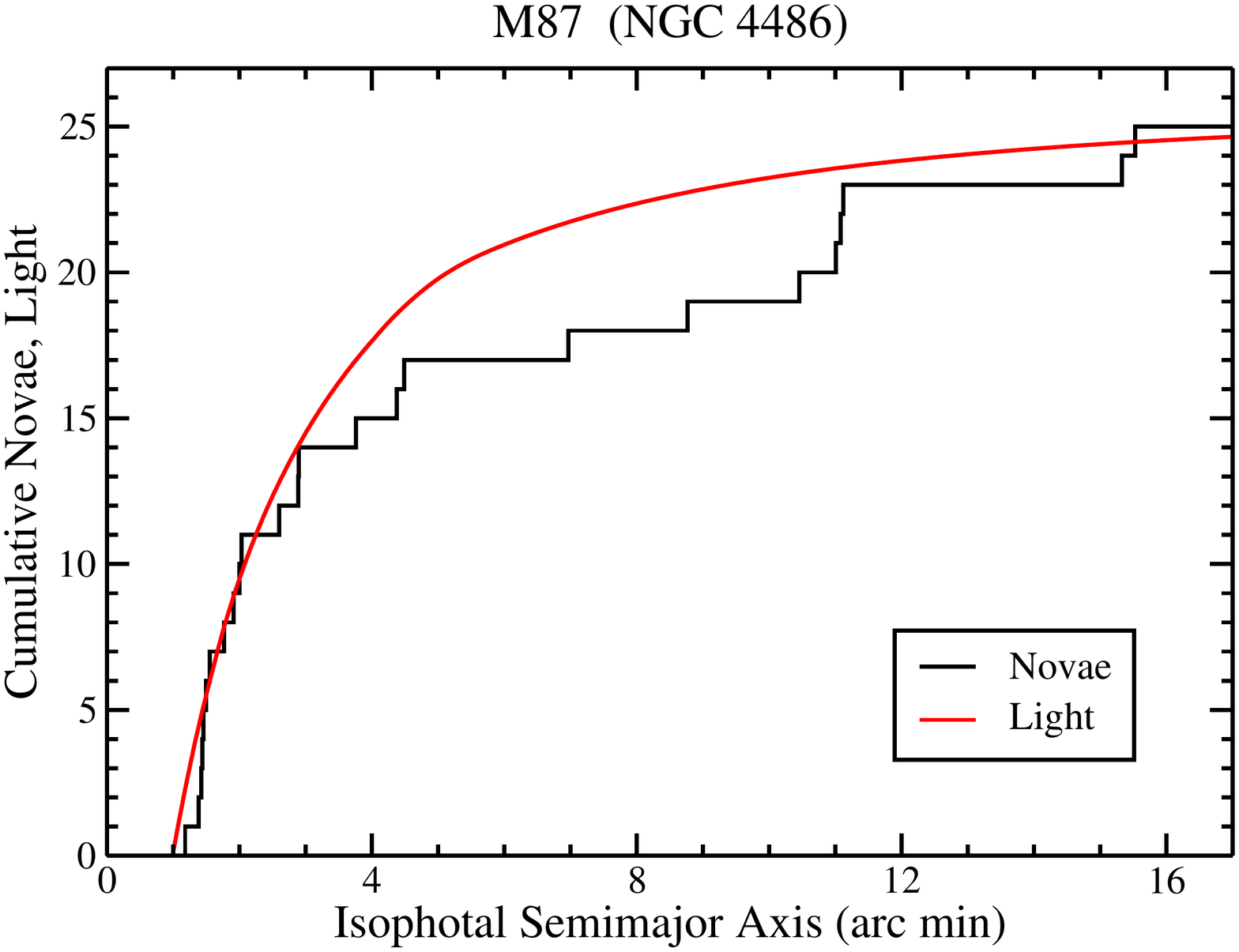} 
\includegraphics[width=0.5\textwidth]{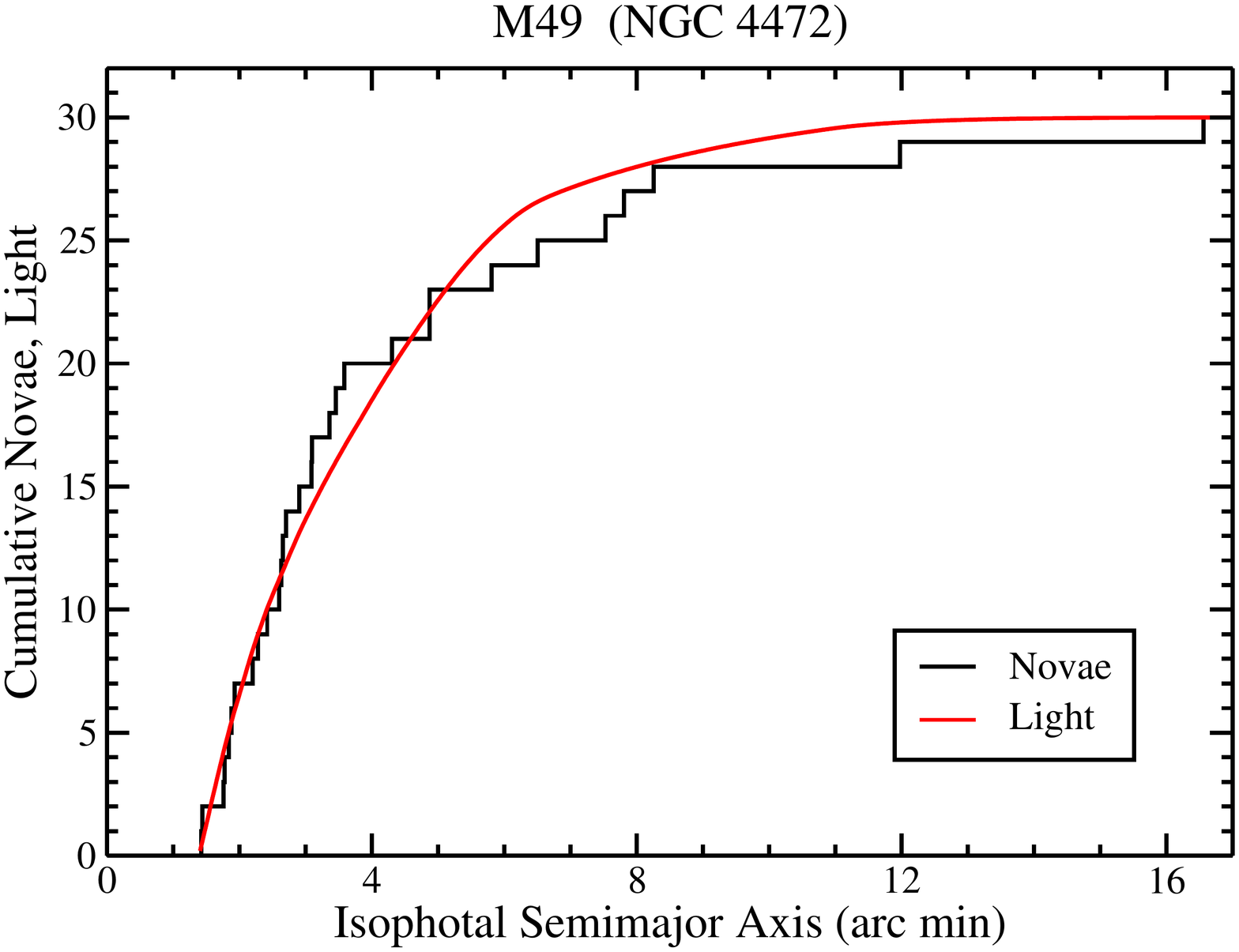} 
\includegraphics[width=0.5\textwidth]{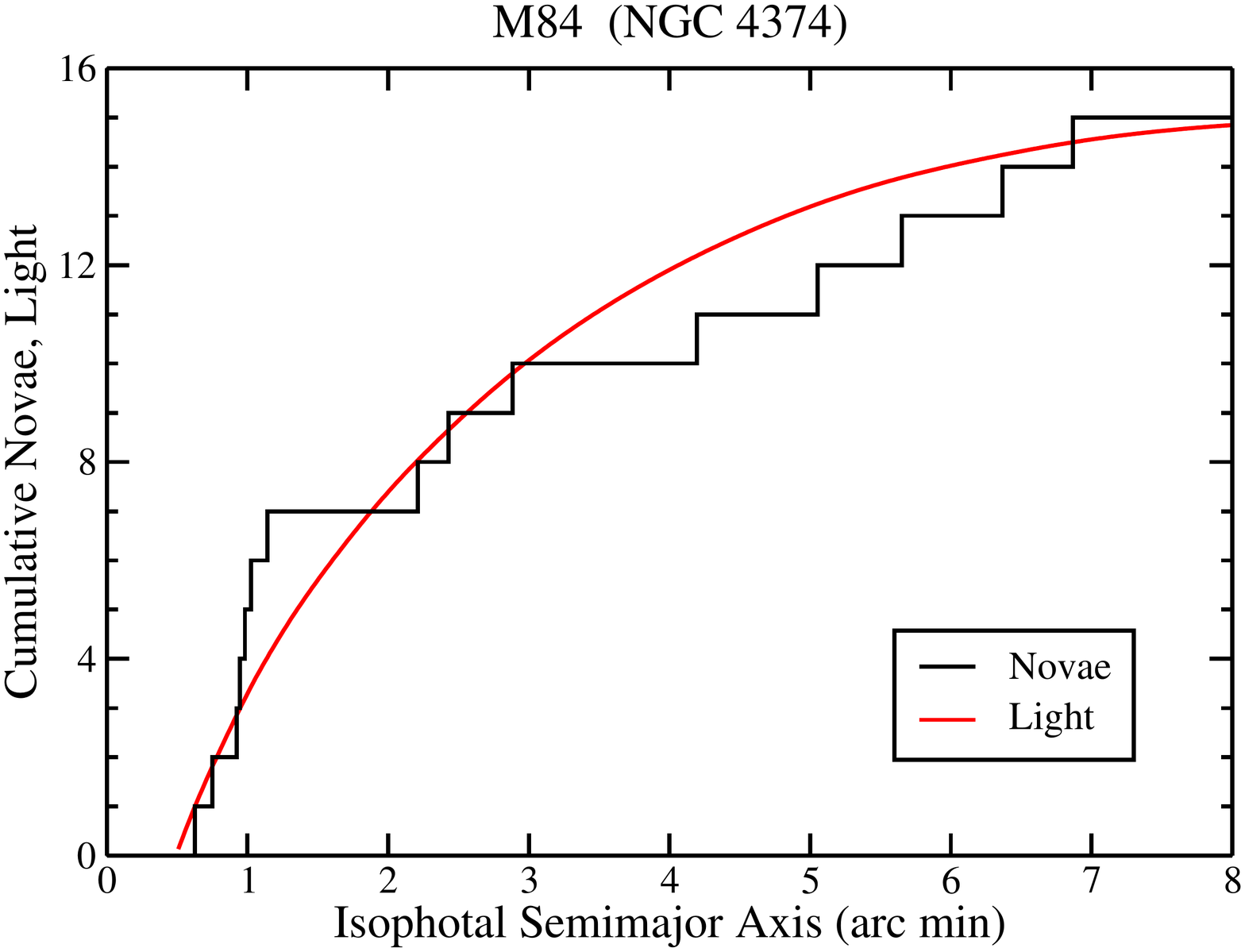} 
\caption{The cumulative distributions of the novae detected in each galaxy in our survey plotted against the cumulative distribution of the background light.
The inner value of $a$ was set to assure the sample was
was complete to the level of the faintest nova. The adopted inner radii
of $a=1.0'$, $a=1.4'$, and $a=0.5'$, for M87, M49, and M84,
resulted in the omission of 2, 7, and 4 novae from the full nova samples,
respectively.
KS statistics for the three fits indicate that the novae
follow the background light in M87, M49 and M84
with 33\%, 66\%, and 56\% probability.
\label{fig2}}

\end{figure}

Equatorial coordinates, the H$\alpha$
magnitude, and the isophotal semi-major axis, $a$,
are given in Table~\ref{tab3}, \ref{tab4}, and \ref{tab5} for the three galaxies, respectively.

\subsection{Spatial Distributions, Completeness, and Galaxy Membership}

The spatial positions of the novae detected in each of
the three galaxies are shown in
Figure~\ref{fig1}. Given
that the Virgo galaxies are members of a rich cluster, it is possible
that some fraction of the novae, especially those seen at a large distance
from their putative host galaxy, may in fact be associated with nearby dwarf
galaxies, or possibly with intracluster novae \citep[e.g., see][]{nei05,sha06}.
To pursue this possibility further, and to
explore how the nova populations are distributed in each of
their host galaxies, in Figure~\ref{fig2} we have compared
the radial cumulative distribution of the novae
with the cumulative distribution of the galaxy's
background light using the $g$-band photometry
of \citet{coh86} for M87 and M49, and the $B$-band photometry
of \citet{mic85} for M84. The different bandpass used for the
M84 photometry should not affect the analysis significantly
given that color gradients in the Virgo ellipticals are not
appreciable \citep[e.g., see][]{coh86}.
In each case the inner value of the isophotal semi-major axis, $a_{in}$,
was set to be the minimum value at which
novae could be detected.
Artificial star tests discussed below in section 4.2
show that our nova samples are complete to the limiting magnitude
beyond $a_{in}=1'$ for M87, $a_{in}=1.4'$ for M49, and $a_{in}=0.5'$ for M84.
Since our surveys were always incomplete within $0.5'$ of the center
of each galaxy, we cannot address the possibility that the nova
density is enhanced near the nuclei of the galaxies, as was suggested by
\citet{mad07} in their {\it HST\/} study of novae in the inner
regions of M87, and by \citet{fer03} for M49.

If the number density of nova progenitor binaries is simply proportional
to the number density of stars, we would expect
the cumulative nova distribution as a function of isophotal semi-major axis
to closely trace that of the galactic background light.
Based on an examination of the cumulative distribution plots, including their
Kolmogorov-Smirnoff (KS) statistics, it appears that
the density of novae follows the background light reasonably well.
In the case of M87 there seems to possibly be two separate populations present.
The M87 distribution is smooth from the central region to $a\sim10'$,
then there is a gap in the distribution followed by what appears to
be an increase in the nova density at $a\grtsim10'$.
A visual inspection of the M87 field reveals that within the
region of this outer population several small neighboring galaxies
are being included in the survey, such as NGC 4478 which by itself
is $\sim10\%$ the luminosity of M87. One of the novae in the outer
population lies within $\sim2.7'$ of NGC 4478. There is also evidence
for an extended halo surrounding M87 \citep[e.g., see][]{kor09}.
Thus is it possible that these outer novae
are not directly associated with M87.

\begin{figure}

\includegraphics[width=0.5\textwidth]{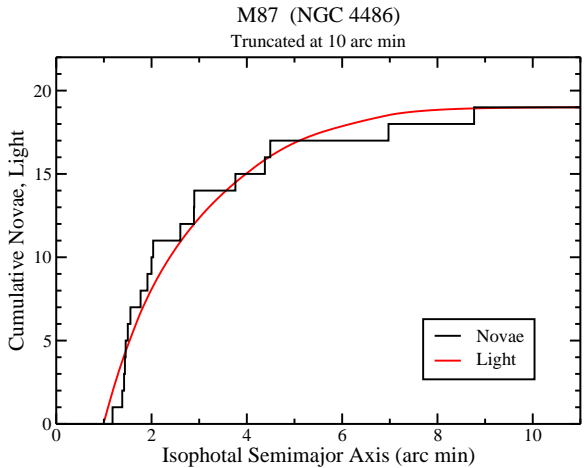} 
\caption{The cumulative distributions of the novae detected in our
 survey of M87 plotted against the cumulative distribution of the
 background light truncated to 10$'$.
 A KS test shows that the distributions match with 81\% probability.
\label{fig3}}

\end{figure}

Given the relatively poor fit of the nova distribution to the cumulative light
in the case of M87
we have considered a truncated sample of novae with $a<10'$
consisting of 21 novae for which there is a high probability of
direct association with the galaxy.
The revised cumulative distribution is shown in Figure~\ref{fig3}.
After removing 6 novae with $a>10'$ and 2 novae with $a<1'$
and rescaling the remaining novae to the
background light accordingly, the KS statistic between the background light
and the nova distribution for M87 improves markedly (KS=81\%).

\begin{deluxetable}{lccccr}
\tabletypesize{\scriptsize}
\tablenum{3}
\label{tab3}
\tablewidth{0pt}
\tablecolumns{6}
\tablecaption{M87 Novae}
\tablehead{\colhead{Nova} & \colhead{} & \colhead{$\alpha$} & \colhead{$\delta$} & \colhead{$m_{{\rm H}\alpha}$} & \colhead{$a$} \\
\colhead{N4486} & \colhead{MJD} & \colhead{(J2000)} & \colhead{(J2000)} & \colhead{(mag)} & \colhead{($'$)}}

\startdata

2011-01a &	55590.9 &		12:30:26.4&		12:24:50&		22.8&		6.97\\
2011-01b &	\dots & 		12:30:36.9& 	12:24:05&		22.2&		3.76\\
2011-01c & 	\dots &		12:30:43.5&		12:22:59&		22.0&		1.77\\
2011-01d & 	\dots &		12:30:44.3&		12:24:53&		22.4&		1.91\\
2011-01e &	\dots &		12:30:49.1&		12:23:08&		21.4&		0.35\\
2011-01f & 	\dots &		12:30:49.7&		12:22:04&		22.3&		1.44\\
2011-01g &	\dots &		12:31:26.1&		12:21:06&		22.3&		11.12\\
2011-03a &	55646.0 &		12:30:23.8&		12:17:22&		21.2&		11.01\\
2011-03b &	\dots &		12:30:37.0&		12:21:47&		21.7&		4.49\\
2011-03c & 	\dots &		12:30:46.0&		12:19:35&		22.4&		4.38\\
2011-03d &	\dots &		12:30:52.2&		12:22:13&		23.0&		1.42\\
2011-03e &	\dots &		12:30:52.3&		12:21:36&		22.3&		2.00\\
2011-03f & 	\dots &		12:30:56.3&		12:13:19&		22.7&		10.46\\
2012-02a & 	55983.0 &		12:30:44.1&		12:24:01&		23.2&		1.50\\
2012-02b &	\dots &		12:30:44.4&		12:21:18&		22.1&		2.89\\
2012-02c & 	\dots &		12:30:44.5&		12:23:03&		22.8&		1.46\\
2012-02d &  \dots &		12:30:47.3&		12:24:45&		22.0&		1.39\\
2012-02e & 	\dots & 		12:30:51.3&		12:23:47&		21.5&		0.59\\
2012-02f & 	\dots &		12:30:55.0&		12:20:55&		23.2&		2.90\\
2012-02g &	\dots &		12:30:56.2&		12:31:15&		22.9&		8.77\\
2012-03a &56006.9 &12:30:14.1&12:35:50&22.8&15.33\\
2012-03b &	\dots &		12:30:28.9&		12:38:09&		22.4&		15.53\\
2012-03c &	\dots & 		12:30:45.2&		12:23:54&		21.9&		1.18\\
2012-03d\tablenotemark{a} &	\dots &		12:30:51.2&		12:21:59&		22.5&		1.56\\
2012-03e & 	\dots &		12:30:52.7&		12:25:06&		21.0&		2.03\\
2012-03f &	\dots &		12:30:58.5&		12:22:57&		22.9&		2.60\\
2012-03g &	\dots &		12:31:26.0&		12:21:46&		22.7&		11.08\\
 
\enddata

\tablecomments{The units for right ascension are hours, minutes, seconds.  The units for declination are degrees, arcminutes, arcseconds.  Apparent magnitudes are approximate only.  The parameter $a$ is the isophotal semi-major axis
of the nova.}
\tablenotetext{a}{This nova was detected coincident with GC M87:[PJB2009] 1667.}

\end{deluxetable}

\begin{deluxetable}{lccccr}
\tabletypesize{\small}
\tablenum{4}
\label{tab4}
\tablecaption{M49 Novae}
\tablewidth{0pt}
\tablecolumns{6}
\tablehead{\colhead{Nova} & \colhead{} & \colhead{$\alpha$} & \colhead{$\delta$} & \colhead{$m_{{\rm H}\alpha}$} & \colhead{$a$} \\
\colhead{N4472} & \colhead{MJD} & \colhead{(J2000.0)} & \colhead{(J2000.0)} & \colhead{(mag)} & \colhead{($'$)}}

\startdata

2011-02a&	55593.9&		12:29:42.5&		07:57:58&			22.2&		2.63\\
2011-02b&	\dots&		12:29:43.4&		07:59:17&			21.9&		1.37\\
2011-02c&	\dots&		12:29:46.2&      	08:01:25&			22.7&		1.44\\
2011-02d&	\dots&		12:29:47.7&      	08:01:44&  			22.9&		1.84\\
2011-02e&	\dots&		12:30:08.4&		07:58:17&			21.4&		6.50\\
2011-03a& 	55650.0&		12:29:30.7&		08:06:24&			21.5&		7.53\\
2011-03b&	\dots&		12:29:34.7&		07:57:30&			22.3&		4.87\\
2011-03c&	\dots&		12:29:37.7&		08:00:15&			21.2&		2.61\\
2011-03d& 	\dots&		12:29:37.9&		07:59.04&			22.3&		2.91\\
2011-03e& 	\dots&       	12:29:42.1&		08:02:54&			21.4&		3.09\\
2011-03f&	\dots&		12:29:46.4&		08:04:41&			23.3&		4.87\\
2011-03g&	\dots&		12:29:47.2&		07:59:30&			21.7&		0.55\\
2011-03h&	\dots&		12:29:48.1&		07:59:32&			21.2&		0.62\\
2011-03i&	\dots&		12:29:50.3&		07:59:10&			22.9&		1.29\\
2011-03j&	\dots&		12:29:52.5&		07:58:58&			23.2&		1.89\\
2011-03k&	\dots&		12:29:52.7&		08:01:46&			23.2&		2.70\\
2011-03l&	\dots&		12:29:53.2&		08:00:09&			22.6&		1.93\\
2012-02a&	55983.1&		12:29:07.3&		08:13:10&			22.4&		16.57\\
2012-02b&	\dots&		12:29:36.6&		07:58:50&			21.9&		3.46\\
2012-02c&	\dots&		12:29:44.5&		07:59:21&			22.7&		1.01\\
2012-02d& 	\dots&		12:29:47.4&		07:58:40&			22.2&		1.42\\
2012-02e&	\dots&		12:29:51.1&		08:00:04&			22.0&		1.33\\
2012-02f&	\dots&		12:29:56.9&		07:54:48&			22.5&		5.81\\
2012-02g&	\dots&		12:30:12.7&		07:55:59&			22.7&		8.26\\
2012-03a&	56007.0&		12:29:18.4&		08:09:34&			23.5&		11.97\\
2012-03b&	\dots&		12:29:44.4&		08:02:36&			22.6&		2.66\\
2012-03c&   \dots&		12:29:44.7&		08:03:02&			22.5&		3.10\\
2012-03d&	\dots&		12:29:46.0&		08:00:45&			22.6&		0.74\\
2012-03e&	\dots&		12:29:47.9&		08:01:39&			22.0&		1.78\\ 
2012-03f&	\dots&		12:29:49.4&		08:02:08&			21.5&		2.42\\
2012-03g&	\dots&		12:29:50.9&		07:58:38&			22.6&		1.76\\
2012-03h&	\dots&		12:29:53.6&      	08:00:51&  			23.5&		2.28\\
2012-03i&	\dots&       	12:29:54.0&		08:00:23&			22.7&		2.20\\
2012-03j&	\dots&		12:29:57.0&		07:58:08&			21.8&		3.36\\
2012-03k&	\dots&		12:29:57.5&		07:56:46&			21.3&		4.30\\
2012-03l&	\dots&		12:29:58.2&		07:58:23&			21.6&		3.58\\ 
2012-03m&	\dots&		12:30:06.1&	 	08:04:02& 	 		21.9&		8.81\\

\enddata

\tablecomments{see notes for Table~3}

\end{deluxetable}

\begin{deluxetable}{lccccr}
\tabletypesize{\small}
\tablenum{5}
\label{tab5}
\tablecaption{M84 Novae}
\tablewidth{0pt}
\tablecolumns{6}
\tablehead{
\colhead{Nova} & \colhead{} & \colhead{$\alpha$} & \colhead{$\delta$} & \colhead{$m_{{\rm H}\alpha}$} & \colhead{$a$} \\
\colhead{N4374} & \colhead{MJD} & \colhead{(J2000.0)} & \colhead{(J2000.0)} & \colhead{(mag)} & \colhead{($'$)}}

\startdata

2011-02a& 	55594.9&          	12:24:40.8&   	12:53:12 	&		23.4& 	5.65\\
2011-02b&	\dots&		12:24:54.3&		12:52:38	&		23.4&		2.43\\
2011-02c&	\dots&		12:25:07.1&		12:53:28	&		21.8&		0.92\\
2011-04a&	55652.8&		12:24:47.6&		12:52:07	&		22.9&		4.19\\
2011-04b\tablenotemark{a}&	\dots&		12:25:02.2&		12:53:43	&		22.9&		0.63\\
2011-04c&	\dots&		12:25:03.5&		12:52:18	&		23.0&		0.98\\
2011-04d&	\dots&		12:25:15.1&		12:52:28	&		21.8&		2.88\\
2012-02a&	55984.1&		12:25:00.3&		12:52:33	&		23.7&		1.14\\
2012-02b&	\dots&		12:25:00.7&		12:53:49	&		23.0&		0.95\\
2012-02c&	\dots&		12:25:01.0&		12:55:15	&		23.2&		2.21\\
2012-02d&	\dots&		12:25:01.8&		12:53:48	&		22.8&		0.75\\
2012-02e&	\dots&		12:25:02.6&		12:53:19	&	    	22.3&    	0.29\\
2012-02f&	\dots&		12:25:02.6&		12:53:32	&	    	22.9&    	0.42\\
2012-02g&	\dots&		12:25:19.6&		12:57:48	&		22.7&		6.37\\
2012-03a&	56007.9&		12:24:47.0&		12:58:38	&		23.0&		6.87\\
2012-03b&	\dots&		12:25:04.2&		12:52:53	&		21.9&		0.37\\
2012-03c&	\dots&		12:25:05.1&		12:53:22	&	    	22.5&    	0.42\\
2012-03d&	\dots&		12:25:06.2&		12:57:57	&		22.0&		5.05\\
2012-03e&	\dots&		12:25:07.8&		12:53:09	&		23.3&		1.03\\

\enddata

\tablecomments{see notes for Table~3.}
\tablenotetext{a}{This nova was detected coincident with a GC, NGC 4374:[CLW2011] 033.}

\end{deluxetable}

\section{Results and Discussion}

\subsection{Relative Nova Rates}

As described earlier, the homogeneous nature of our survey, where we
monitored each galaxy in a consistent fashion,
allows us to estimate the relative nova rates simply
by comparing the relative numbers of novae detected.
We have found that the relative number of novae discovered in each galaxy
(27 in M87, 37 in M49, and 19 in M84) is in good agreement with
that predicted under the null hypothesis that the nova rate scales
directly with galaxy luminosity. This is true whether we consider the
entire population of novae discovered in the extended field surrounding each
galaxy, or, in the case of M87,
whether we consider the more restrictive sample (21 novae) based
on the best match of the cumulative nova distribution with the background
light. Thus, based solely on the relative numbers of novae
discovered in the three galaxies, we find no evidence that the GC
specific frequency plays a significant role in the nova
production in the host galaxies.

\subsection{Absolute Nova Rates}

Although our program was designed to compare relative
nova rates in the three Virgo ellipticals, it is also possible to
use our data to estimate absolute nova rates in the galaxies as well.
Our procedure follows that employed in earlier studies where we
used a Monte Carlo approach to estimate the nova rates
\citep[e.g.,][]{fra12}. In this approach, for a given
assumed intrinsic nova rate, $R$,
we compute the number of novae that we can expect to observe, $N_{\rm obs}(R)$,
given assumed properties of the nova population (peak brightnesses, and
fade rates), our observing cadence, the distance to the galaxy in question
and the limiting magnitude of our survey at each epoch, $i$, as a function
of position in the galaxy.
For a range of plausible values of $R$, the simulation constructs a
set of model H$\alpha$ light curves by randomly selecting peak
magnitudes and decay rates from a sample of actual H$\alpha$ nova
light curves from M31 \citep{sha01} and M81 \citep{nei04}.
Given the dates of our observations (see Table~\ref{tab2}) and the distance
modulus to the galaxy (see Table~\ref{tab1}), an observed nova luminosity
function, $n_i(m,R)$, for each epoch is calculated.
Following \citet{fra12},
the number of novae we can expect to observe as predicted by our Monte Carlo
analysis is then given by:

\begin{equation}
N_{\rm MC}(R) = \sum_{i}\sum_{m}{C(m+\Delta m_i)~n_i(m,R)},
\end{equation}

\noindent
where $C(m)$ is the
completeness as a function of apparent magnitude, which
accounts for the variation in limiting magnitude across the galaxy.

\begin{figure}

\includegraphics[width=0.5\textwidth]{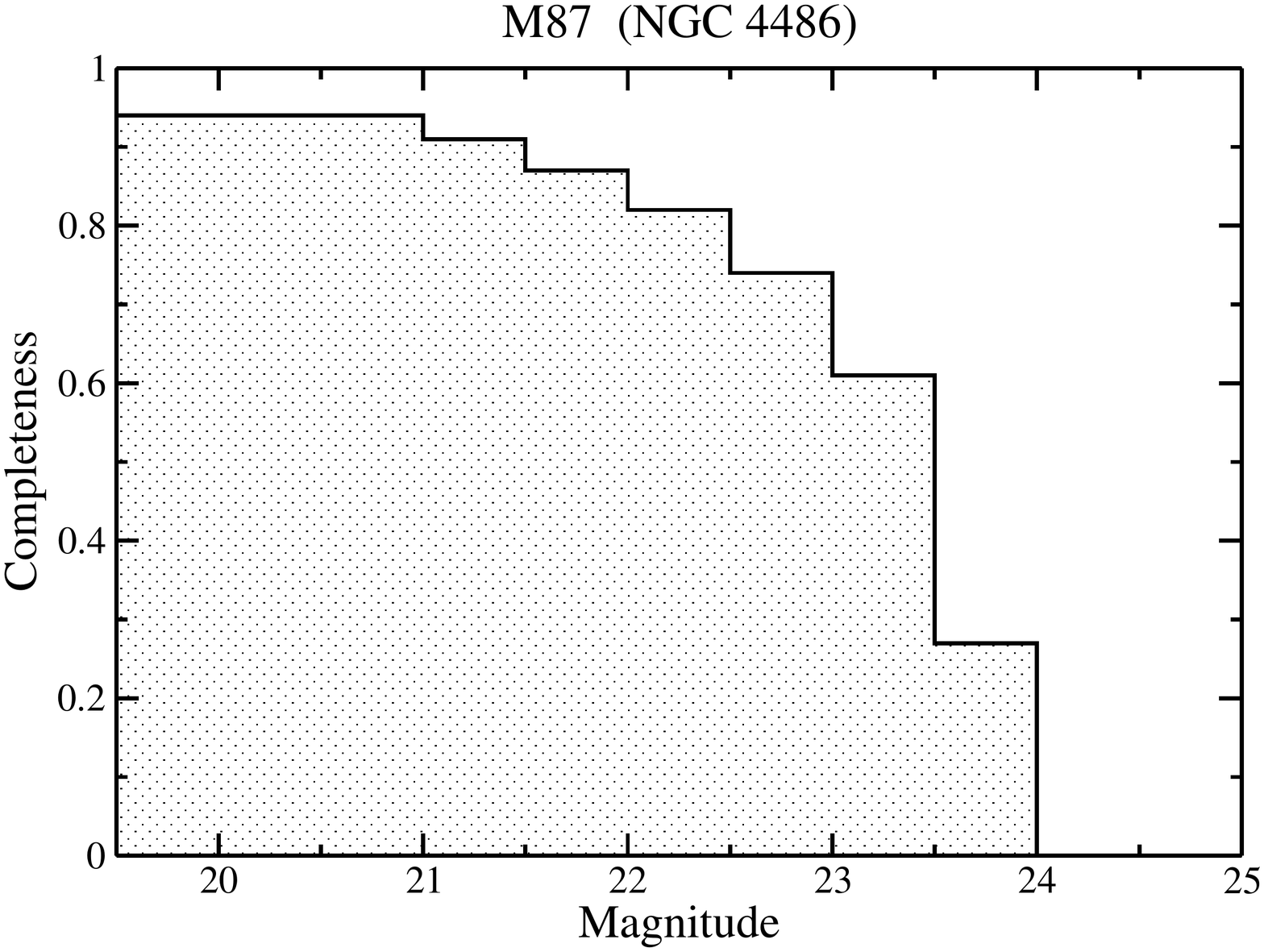} 
\includegraphics[width=0.5\textwidth]{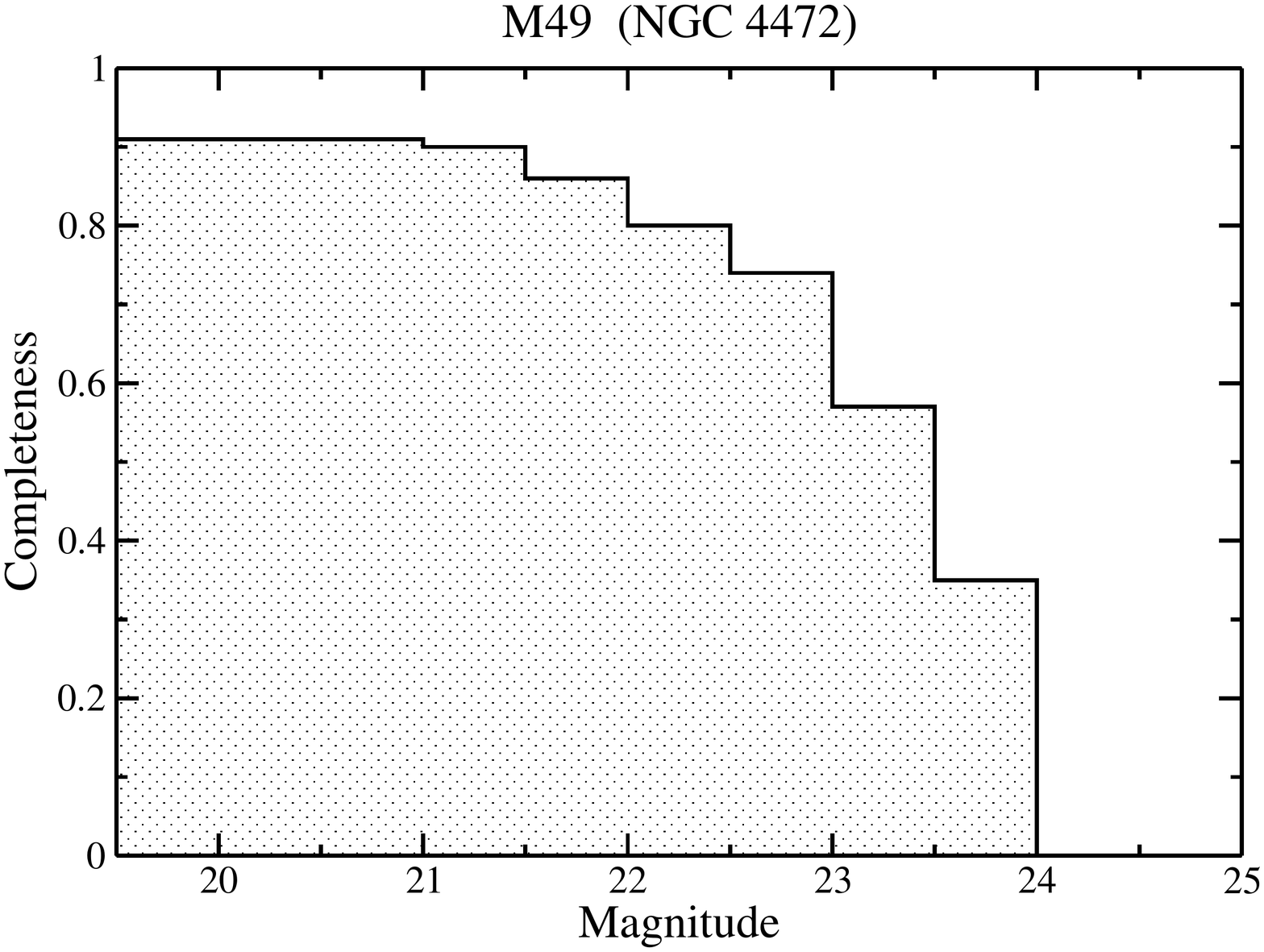} 
\includegraphics[width=0.5\textwidth]{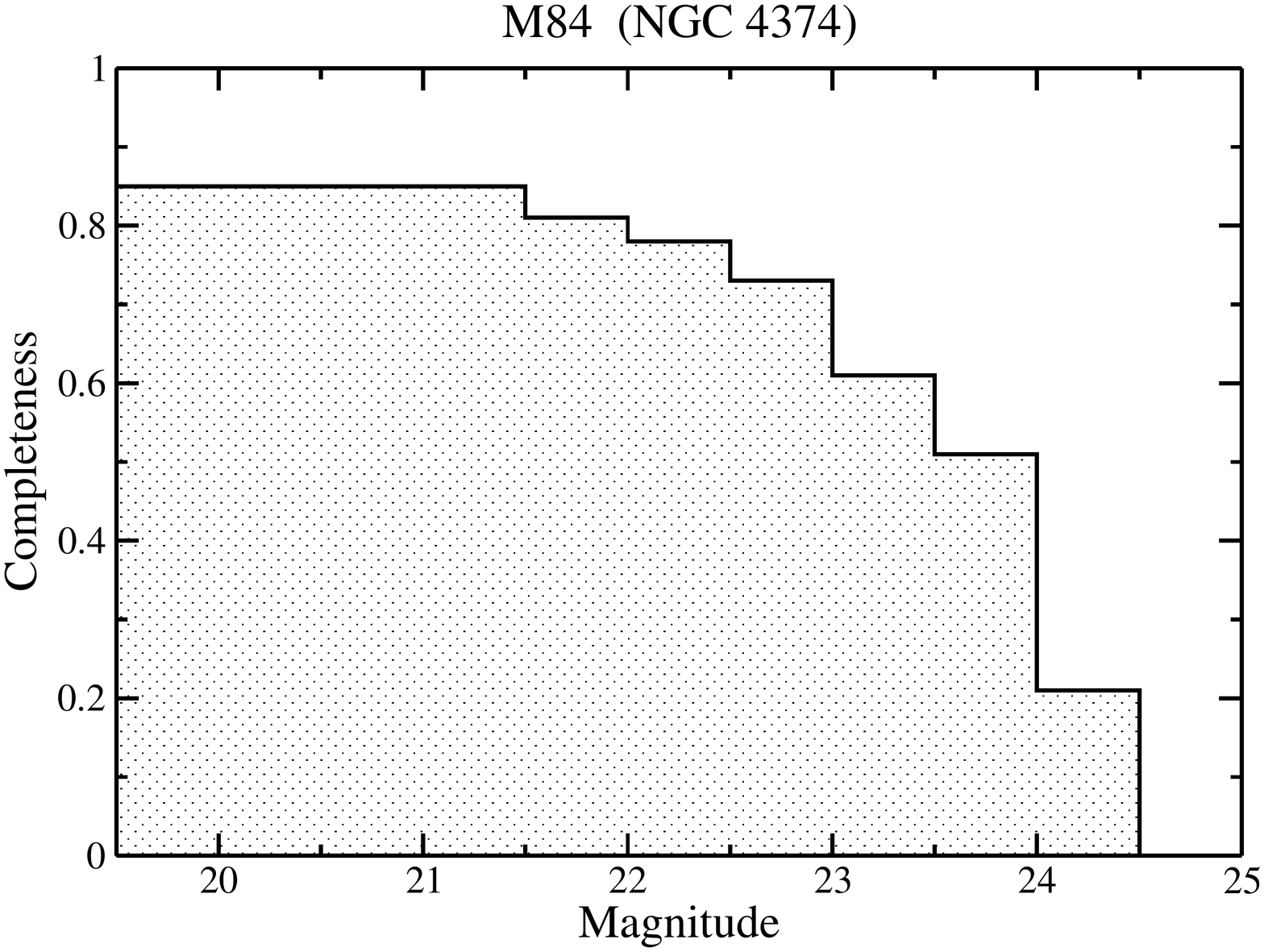} 
\caption{The completeness functions, $C(m)$,
showing the fraction of artificial novae recovered in our artificial
nova simulations for the different galaxies as a function of apparent magnitude.
\label{fig4}}

\end{figure}

\begin{figure}

\includegraphics[scale=0.3]{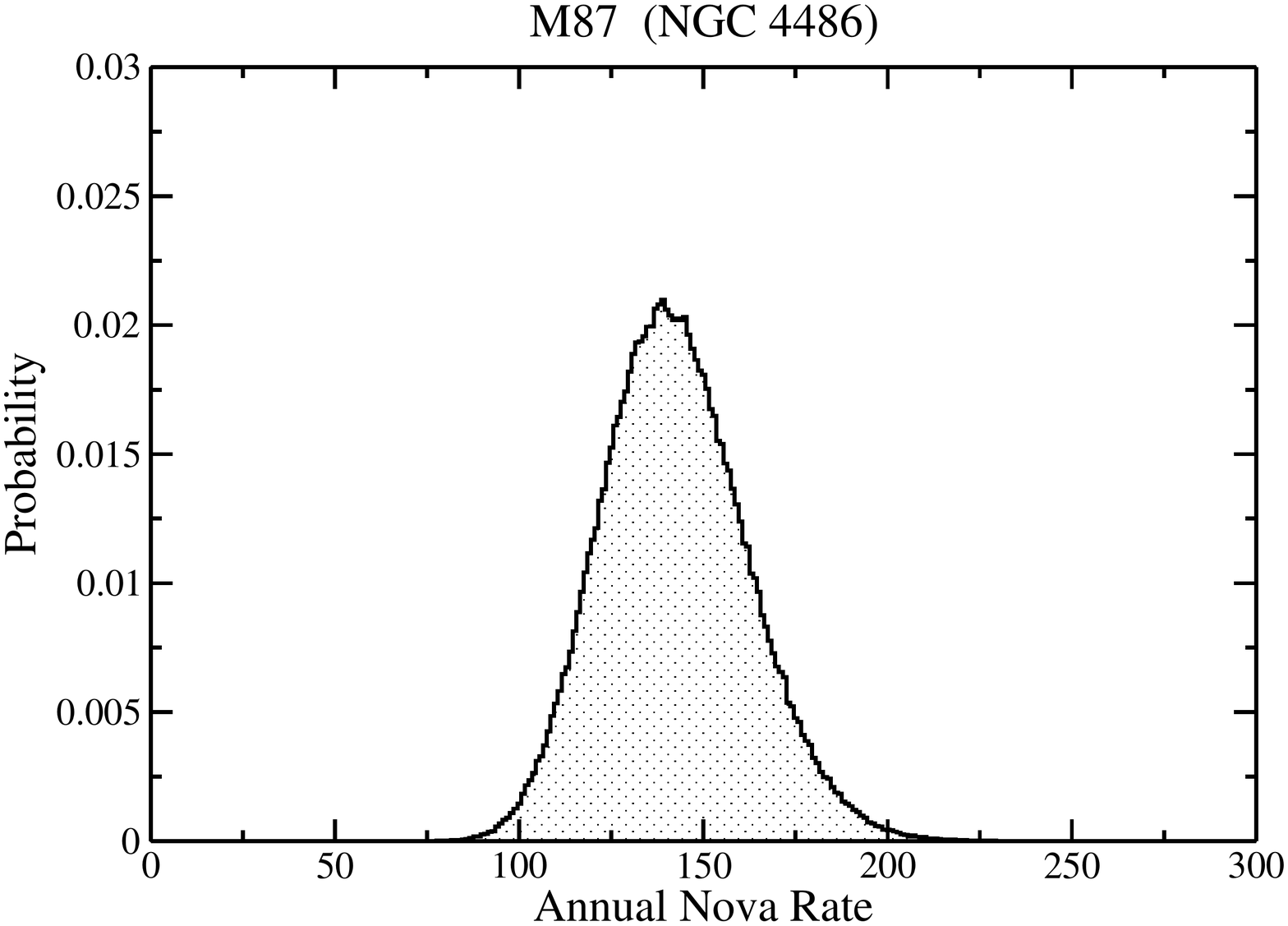} 
\includegraphics[scale=0.3]{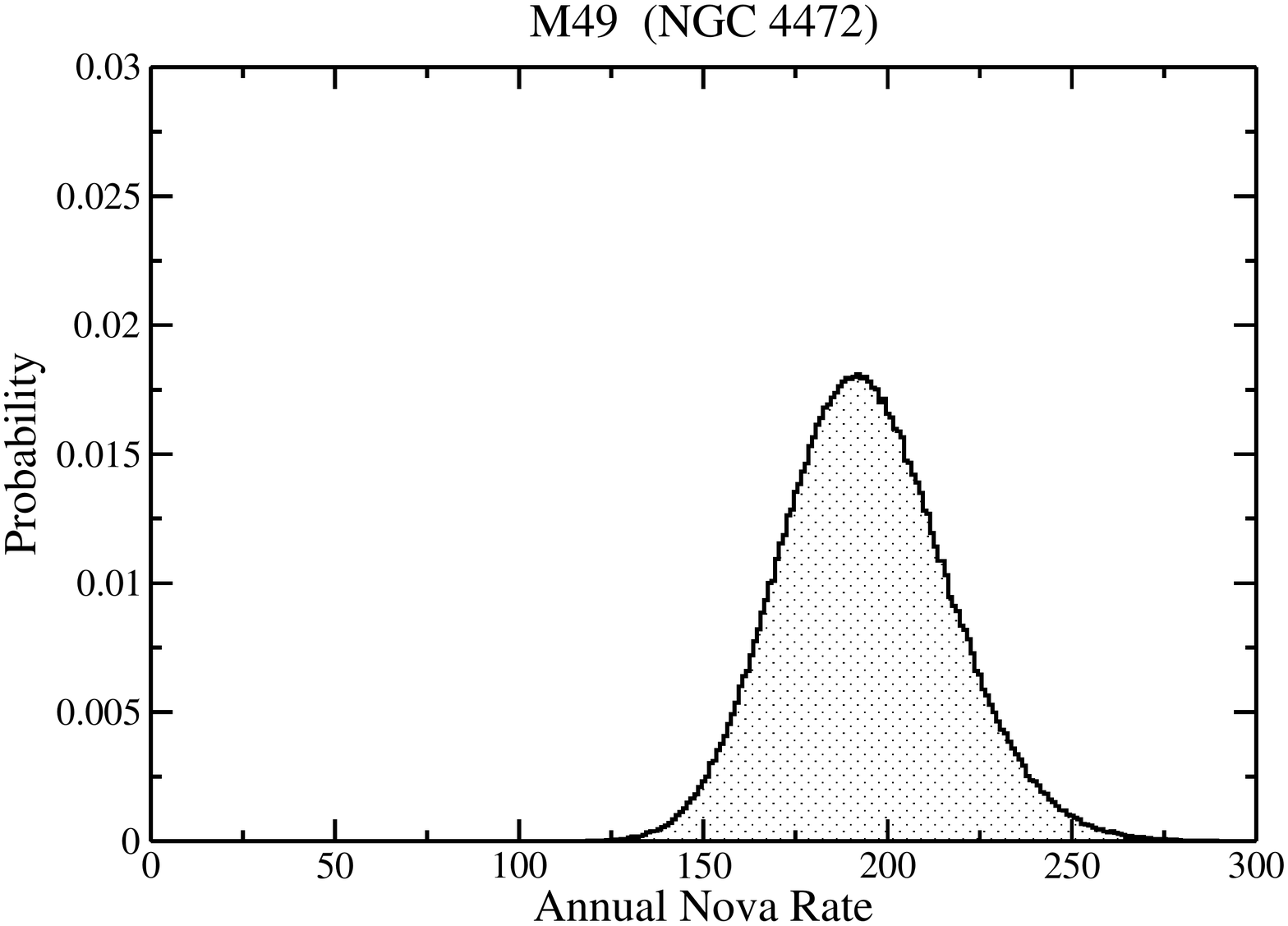} 
\includegraphics[scale=0.3]{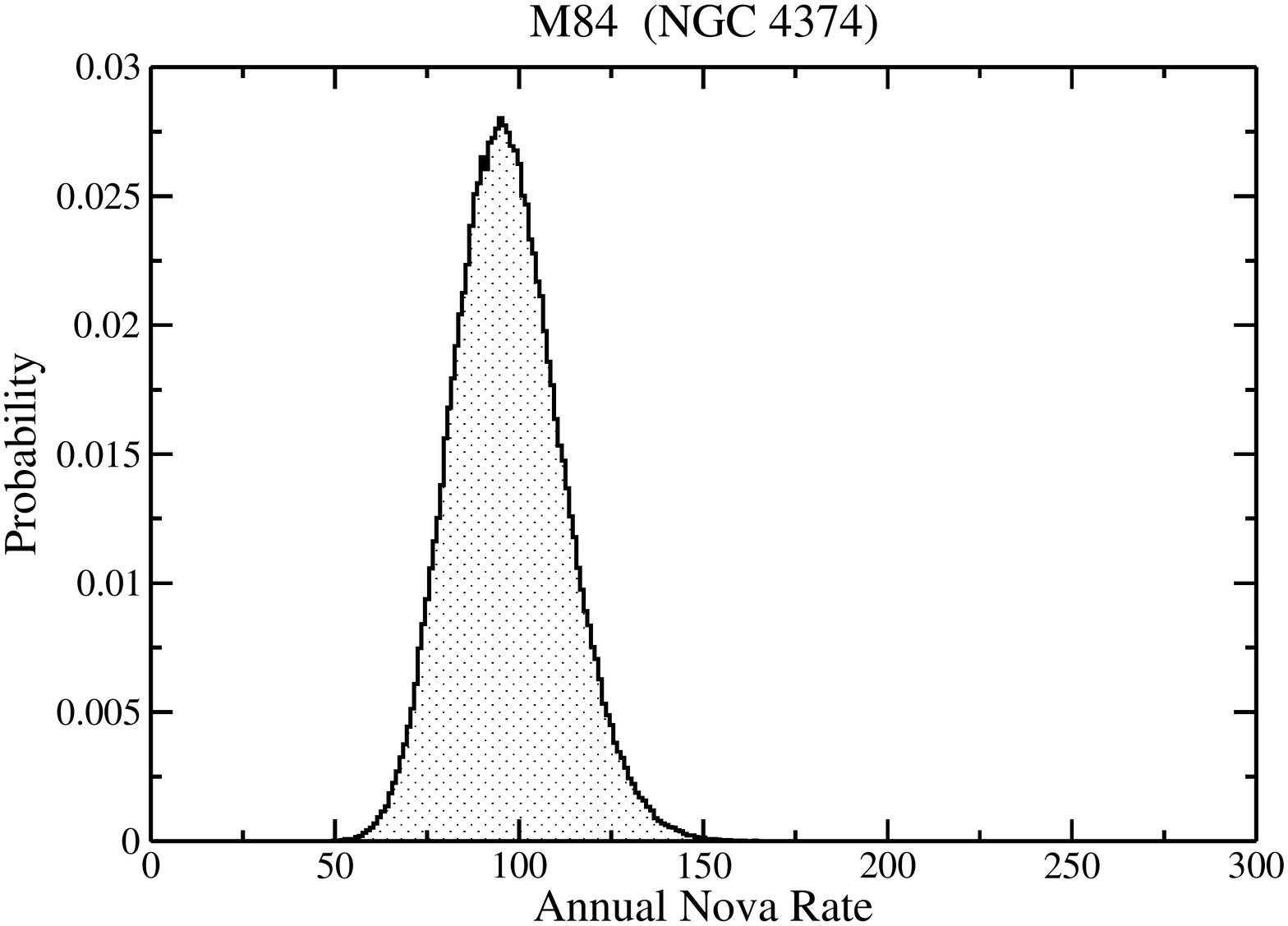} 
\caption{The probability distributions of the
annual nova rates, $R$. The peak of each distribution represents
the most likely nova rate, with 1$\sigma$ errors determined assuming
that the distribution is Bi-Gaussian (see text).
\label{fig5}}

\end{figure}

The completeness function is
estimated through artificial star simulations using the IRAF
task {\tt ADDSTAR}.
Artificial novae are generated in several
magnitude bins spanning the range of nova magnitudes observed.
For each bin, the artificial novae
are distributed throughout the galaxy image, with a surface
density that follows
the background light. Then, employing the same search technique
used to discover the real novae, we determine the fraction of
artificial novae that we can recover as a function of
magnitude.
Completeness functions for the three galaxies are shown
in Figure~\ref{fig4}.
Variation in the completeness at different
epochs, which may differ slightly in exposure time and seeing,
is compensated for by estimating an offset,
$\Delta m_i = m_{lim,0} - m_{lim,i}$, between the epoch used in the
artificial star tests and the other epochs.

The most probable nova rate for a given galaxy is
given when the number of novae actually observed, $N_{\rm obs}$, matches the number
of novae predicted by our Monte Carlo analysis, $N_{\rm MC}(R)$.
The Monte Carlo program repeats this procedure 100,000 times and records
the number of times $N_{\rm MC}(R) = N_{\rm obs}$ for each galaxy.
The number of matches as a function of $R$,
normalized to the total number of trials, produces the probability
distribution of nova rates for each galaxy shown in Figure~\ref{fig5}.
The peak of this distribution represents the most likely intrinsic
nova rate, $R$, with $1\sigma$ errors estimated by assuming Bi-Gaussian
fits to the probability distributions \citep{buy72}.
We have also computed a nova rate for M87 based upon the truncated
sample with $a<10'$.
A summary of the nova rates
for each galaxy is presented in Table~\ref{tab6}.
We note that the rates determined for M87 and M49, in particular,
are somewhat higher that those
found previously in studies by \cite{sha00} and \cite{fer03}.
Despite our higher nova rate for M87, our value of $154^{+23}_{-19}$~yr$^{-1}$
for the full sample falls short of the rate of $\sim200-300$~yr$^{-1}$
suggested by \citet{sha02} from early {\it HST\/} data, but is similar to
that found by \citet{miz13}.

In addition to computing nova rates, we have also determined the
$K$-band luminosity-specific nova rate, $\nu_K$ for each galaxy
from the data given in Table~\ref{tab1}. In all cases we find values of $\nu_K$
of $\sim3$ novae per year per $10^{10}$ solar luminosities in the $K$ band.
Our results are somewhat higher, but still
marginally consistent with the findings of \cite{wil04}
and \cite{gut10} who found for a large sample of galaxies
that $\nu_K\approx2\pm1$ independent of Hubble type.
Our $K$-band
luminosity-specific nova rates are presented in Table~\ref{tab6}.

\begin{deluxetable}{lccc}
\tabletypesize{\scriptsize}
\tablenum{6}
\label{tab6}
\tablecaption{Nova Rates}
\tablewidth{0pt}
\tablecolumns{4}
\tablehead{
\colhead{} & \colhead{} & \colhead{}  & \colhead{$\nu_K$} \\
\colhead{Galaxy} & \colhead{$N_{\rm obs}$} & \colhead{$R~(yr^{-1})$} & \colhead{$(10^{10}~L_{\odot,K}\cdot yr)^{-1}$}
}

\startdata

M87                 &		27&	$154^{+23}_{-19}$&      $3.8\pm1.0$	\\
M87$_{T}$\tablenotemark{a}&	21&	$124^{+23}_{-16}$&	$3.1\pm0.8$	\\
M49             &		37&	$189^{+26}_{-22}$& 	$3.4\pm0.6$	\\
M84&	  			19&	$95^{+15}_{-14}$& 	$3.0\pm0.6$	\\
\enddata

\tablenotetext{a}{Only novae interior to an isophotal semimajor axis of 10$'$ have been considered in this analysis.}

\end{deluxetable}

\section{Globular Cluster Novae}

\begin{figure}
\includegraphics[angle=0,width=.236\textwidth]{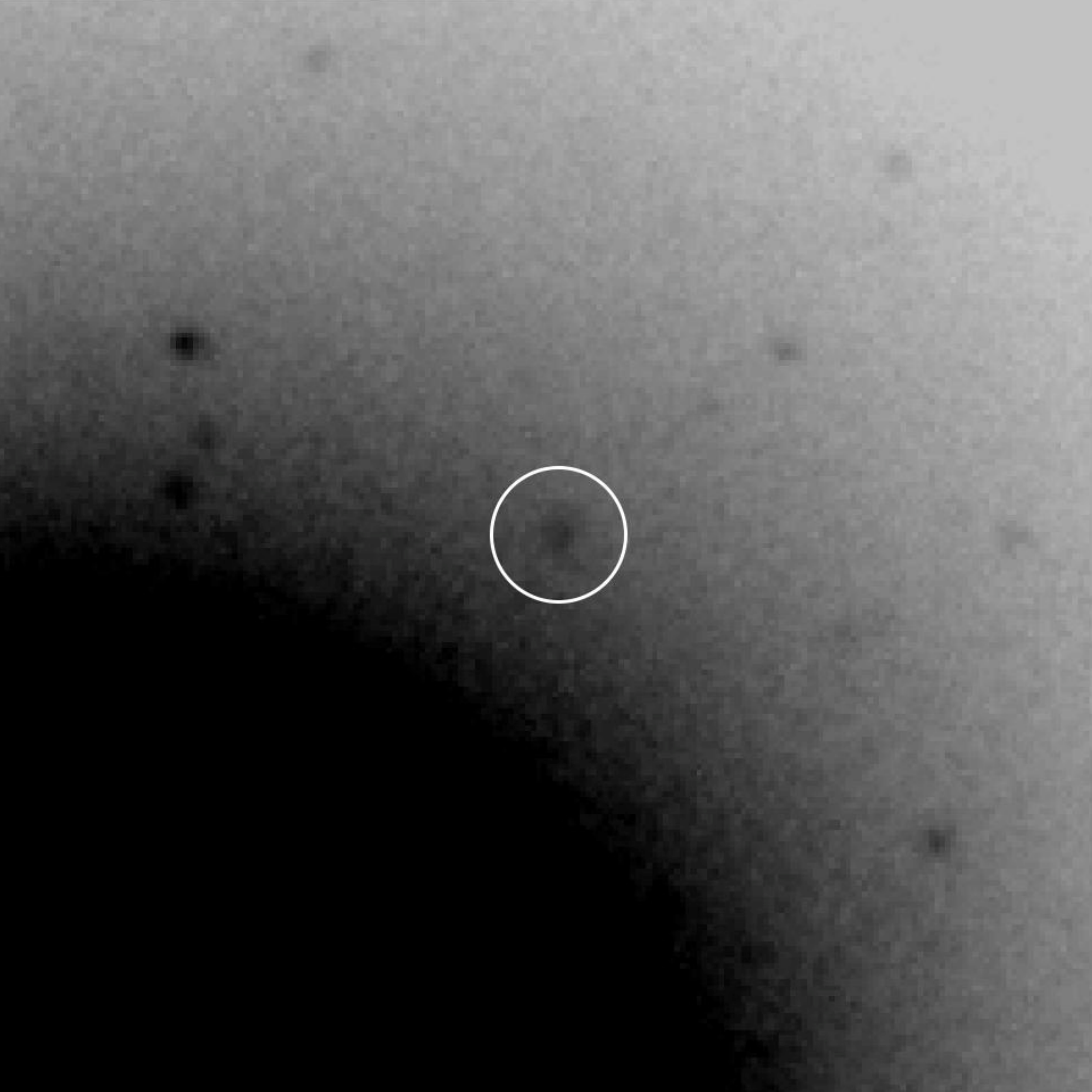}
\includegraphics[angle=0,width=.236\textwidth]{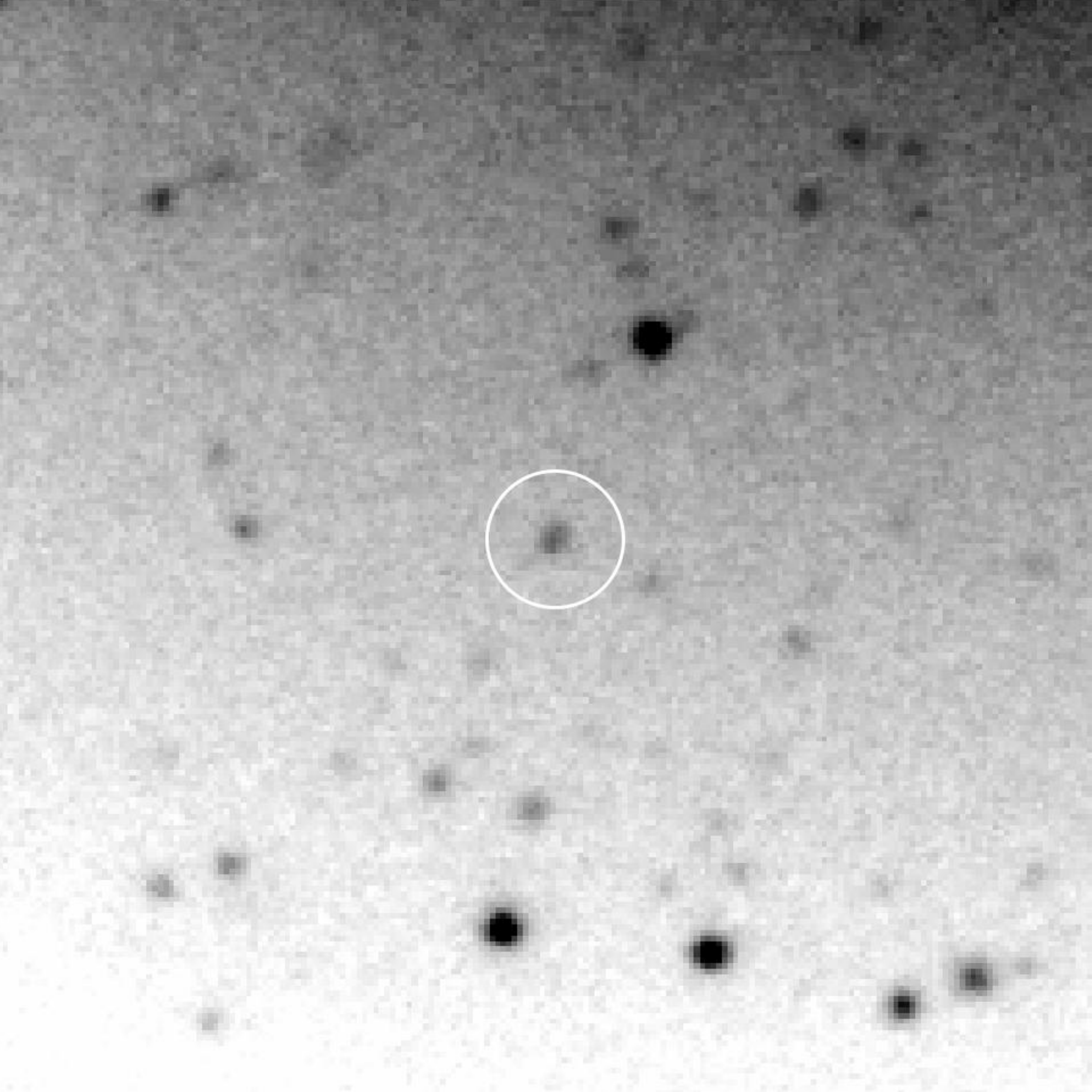}
\includegraphics[angle=0,width=.236\textwidth]{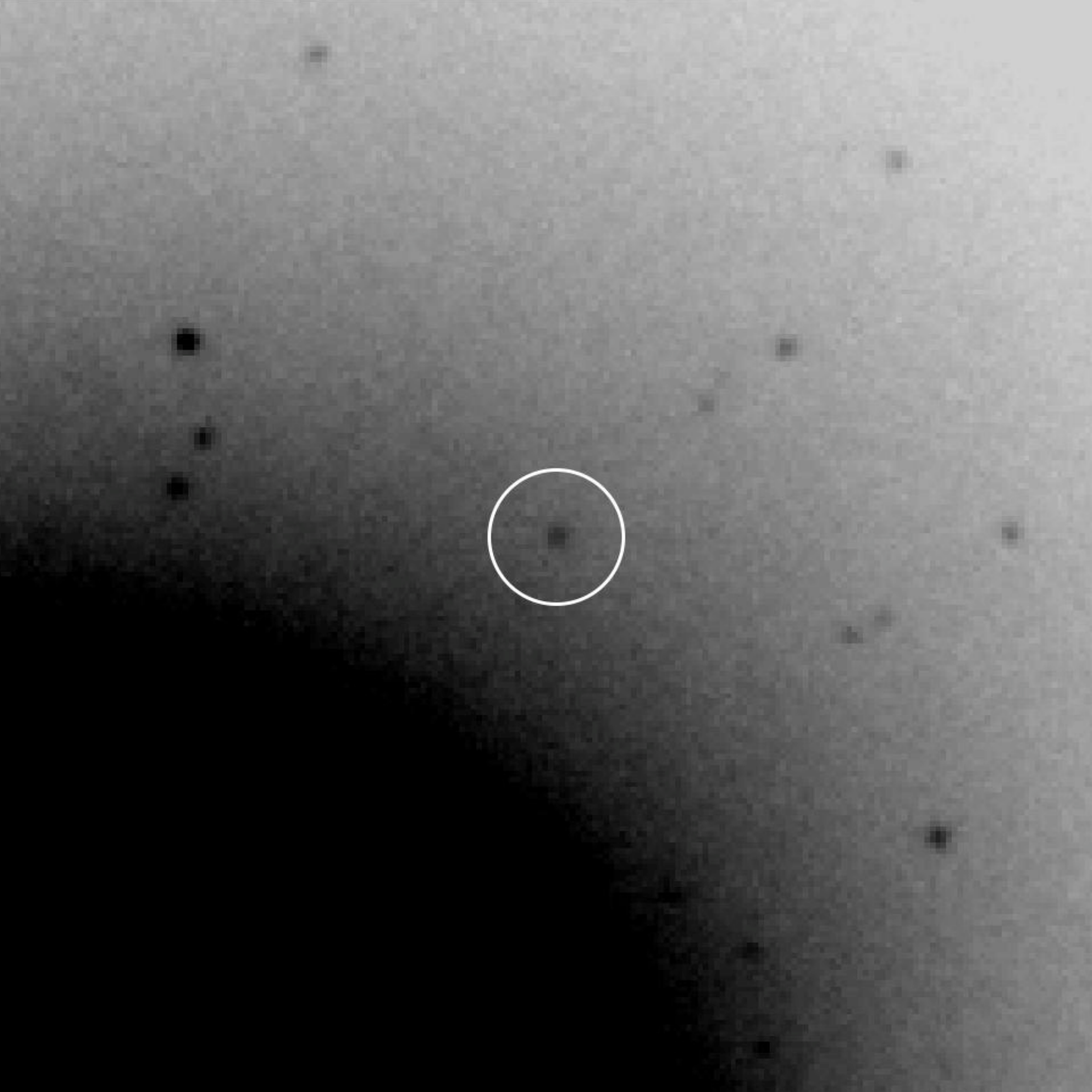}
\includegraphics[angle=0,width=.236\textwidth]{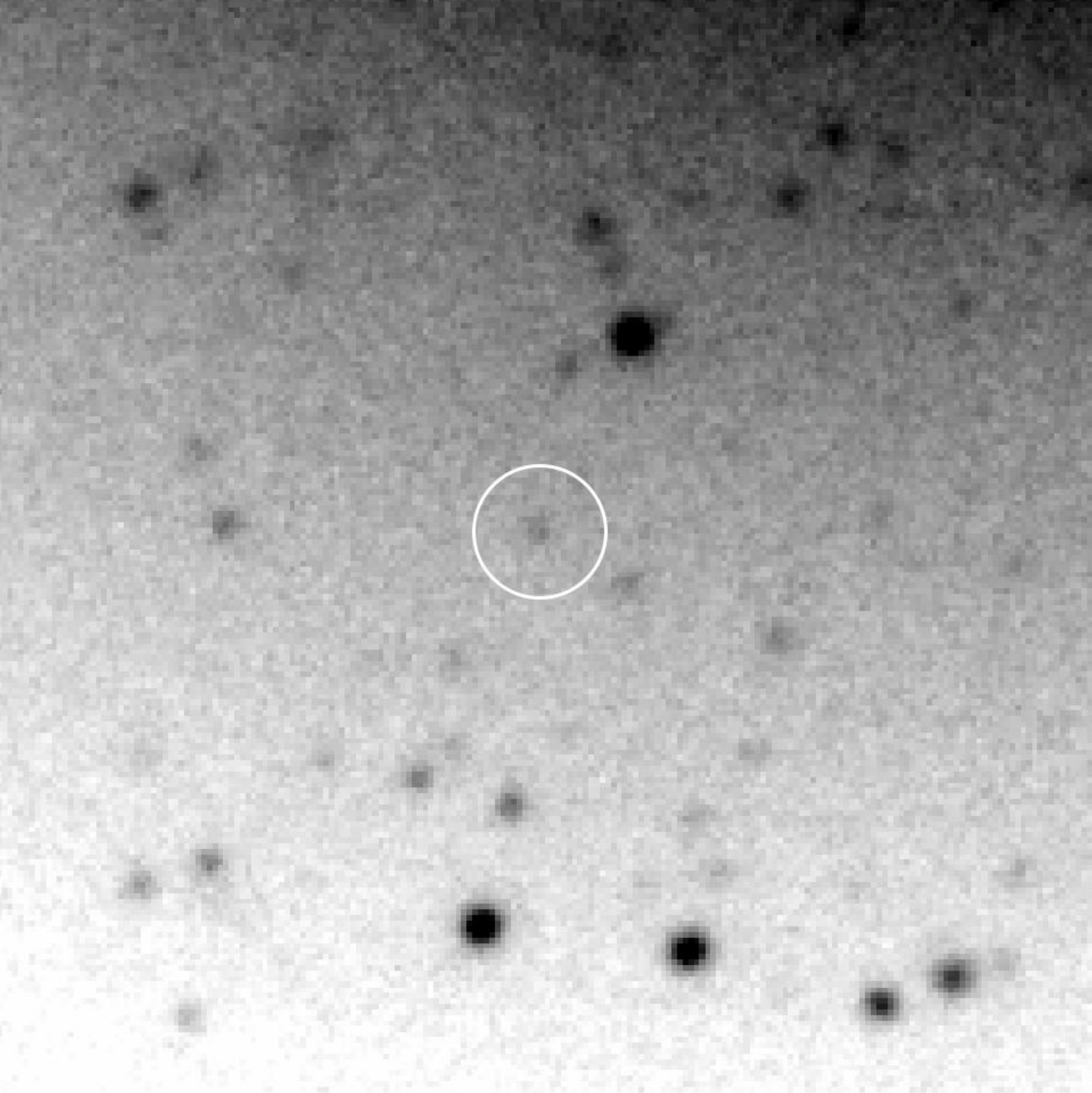}
\includegraphics[angle=0,width=.236\textwidth]{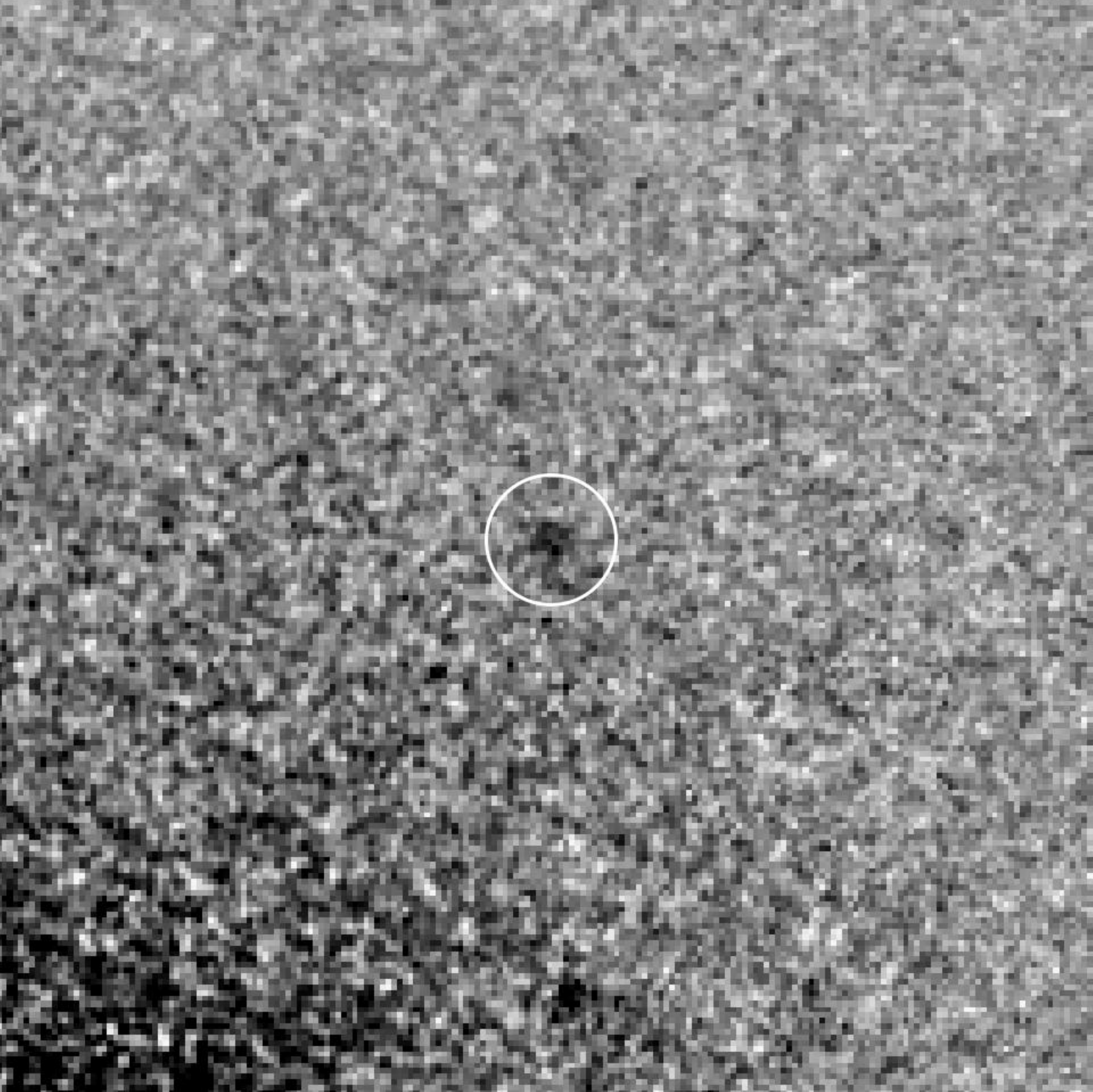}
\includegraphics[angle=0,width=.236\textwidth]{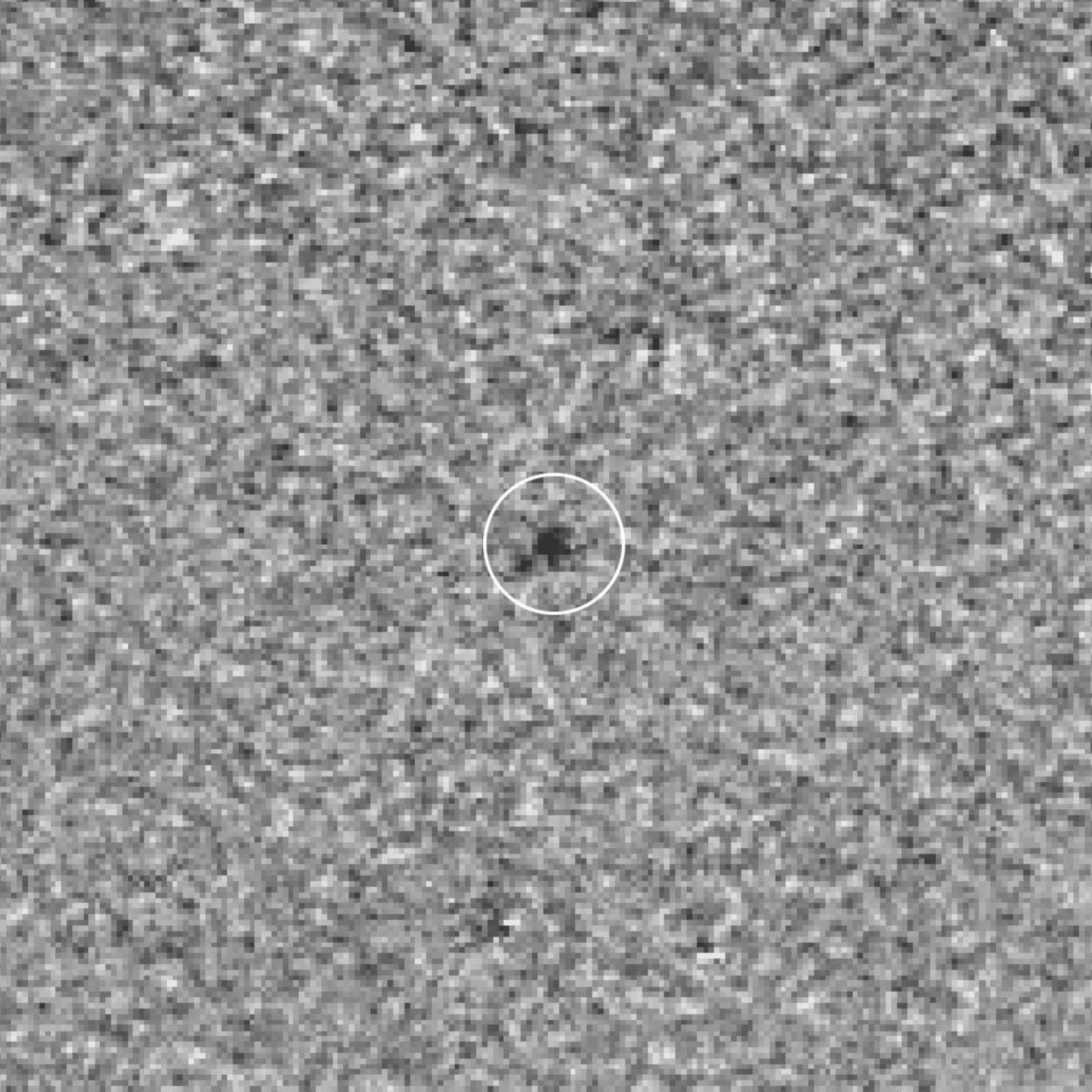}
\caption{Left column: Images of the M84 GC CLW2011-033 from Epoch 2
(with nova, top),
from Epoch 4 (without nova), and their difference. Right column:
Images of the M87 GC PJB2009-1667 from Epoch 4 (with nova, top),
from Epoch 1 (without nova), and their difference.
In both cases residual light in the GCs is detected between epochs
that we attribute to erupting novae.
North is up and East to the left, with a scale of $\sim40''$ on a side.
\label{fig8}}
\end{figure}

Two of the novae discovered in our survey, one in M87 and one in M84,
appear to be spatially coincident with GCs associated
with these galaxies. No GC novae were identified in M49.
The GC nova candidates
were discovered from the detection of residual light in differenced
images at the positions of known GCs.
Instrumental magnitudes were measured for the globulars in question,
as well as for several nearby clusters, for all four epochs.
While the magnitudes of the surrounding clusters remained reasonably constant,
the two GCs thought
to host novae were seen to increase significantly
in brightness in only one epoch.
Figure~\ref{fig8} shows images of the M84 and M87 GC nova fields
from epochs
when the novae were detected, quiescent epochs, and their differences.
We attribute the residual light to erupting novae in these clusters.
Nova magnitudes were estimated by
subtracting the contribution of the host GC determined
from quiescent epochs.
The resulting nova magnitudes given in Tables~\ref{tab3} and~\ref{tab5},
were consistent with those of the field novae, providing further
evidence that the cluster brightenings were indeed the result of novae
erupting from within the cluster.

The PSFs of the GCs with
erupting novae were consistent with those for other GCs,
and were indicative of a single, unresolved light source.
The centroid of the PSF for the GCs with erupting novae
were coincident with their quiescent counterparts
in all epochs to within $0.1''$,
much less than the astrometric error of $1''$.
We are sufficiently confident that these events are in fact GC novae
to include them in our final count.

The M84 GC, CLW2011:033, is a
relatively blue cluster characterized by $g=22.853\pm0.016$ and
$g-z=0.997\pm0.024$ \citep{chi11}.
The cluster appears to be metal poor and relatively massive,
being $\sim1$ mag brighter than the turnover magnitude.
On the other hand the M87 cluster, PJB2009:1667, is characterized
by $V=23.235\pm0.007$ and $V-I=1.183\pm0.010$ \citep{pen09}. This cluster
is redder than the M84 clster, more metal rich, with a flux
that is $\sim0.5$ mag brighter than the turnover magnitude.

Despite predictions that novae may be common in
GCs, relatively few have been reported. There have been just
seven novae identified with GCs: four
in M31 [M31N 2007-06b in Bol~111 \citep{sha07},
CXO~J004345 in Bol~194 \citep{hen09}, M31N 2010-10f in Bol~126
\citep{cao12,hen13}, and possibly an anonymous source
in Bol~383 \citep{pea10a}],
two in the Galaxy [T~Sco (Nova Sco 1860) in M80 \citep{lut60,pog60},
and possibly Nova Oph 1938 in M14 \citep{hog64,mar91}],
and an anonymous source in an M87 cluster \citep{sha04}.
Until now, the latter object was
the only GC nova that has been reported in M87,
with no cluster novae known in either M49 or M84.
Clearly, the paucity of novae detected in GCs to date is in part
a consequence of the difficulty in discovering novae against the
bright background light of the cluster.

The discovery of two GC
novae from our survey provides clear
evidence that GCs do contribute some
fraction of a galaxy's nova system population, and that novae seem to be
enhanced in GCs relative to the field.
We can estimate the GC enhancement by estimating the mass in GCs
in the three galaxies and comparing it with the integrated galaxy masses.
The total number of GCs
is estimated to be 13,200 in M87 \citep{har98}, 5900 in M49 \citep{rho04},
and 1775 in M84 \citep{gom04}.
Using a mean GC mass of $2.4\times10^5$~M$_{\odot}$
appropriate for massive Virgo ellipticals \citep{mcl99},
we estimate that the GC systems
have a combined mass of $\sim5\times10^9$~M$_{\odot}$.
For the galaxy masses, we
adopt $7.6\times10^{11}$~M$_{\odot}$, $8.1\times10^{11}$~M$_{\odot}$, and
$4.0\times10^{11}$~M$_{\odot}$ for M87, M49, and M84,
respectively \citep[][Table 1]{ang97}, yielding
a combined mass of $\sim2.0\times10^{12}$~M$_{\odot}$.
Thus, GCs represent
$\sim2.5\times10^{-3}$ of the total mass in the three galaxies.
Given that we
have detected two GC novae out of a total of 83 novae
discovered in our survey, we estimate that novae erupt
$\sim 10$ times more frequently in GCs than they do in the field.

\citet{sha04} estimated that novae occur in M87 GCs with
a frequency, $f\sim4\times10^{-3}$ novae per GC per year, which is $\sim2$
orders of magnitude higher than what would be expected if novae were not
enhanced relative to the field. In our observations
of M87 we have discovered one GC nova out of 27 total novae. Assuming
M87 has 13,200 GCs and an overall nova rate of $\sim150$~yr$^{-1}$, we estimate
a GC rate of $f\sim4\times10^{-4}$ novae per cluster per year,
or about an order of magnitude less than that estimated
by \citet{sha04} for this galaxy.


We can estimate the significance of the GC enhancement
as follows. Let $p=2.5\times10^{-3}$ be the fraction of a galaxy's mass in GCs.
The probability of observing $N$ or more
GC novae out of a sample of $M$ total novae
assuming {\it no\/} GC enhancement is simply
\begin{equation}
P_{\geq N,M} = \sum_{N=2}^\infty {M! \over N! (M - N)!} \  p^N \, (1 - p)^{M-N}.
\end{equation}
The probability of observing 2 or more GC novae out of 83 novae
is then $P_{\geq 2,83}\simeq0.018$, or $\sim1.8$\%.
In other words, we conclude that novae are enhanced in GCs
with 98.2\% confidence.
Our estimate likely represents a lower limit to the true GC nova
enhancement given the difficulty in identifying novae against the
GC background light. The selection effect is amplified when we consider that
the most luminous GC have the densest cores, and are thus the most likely
to form nova progenitor binaries.

Novae discovered in GCs can provide insights
into the dynamics of the clusters and isolate the
effects of physical parameters such as age and metallicity
in the (primarily) simple stellar populations of GCs.
Low mass X-ray binaries, which are the only other binary systems
that can be studied in Virgo cluster galaxies, provide an
interesting point of reference.
These compact neutron star and white dwarf systems are dynamically enhanced
in GCs \citep{cla75,hil76} by a factor of several hundred,
and show several clear trends. The X-ray binary rate is correlated
with the mass and metallicities of the host GCs.
X-ray binaries are preferentially found the most massive GCs
because the density of GCs increases with mass
leading to a very strong correlation between the dynamical interaction rate,
cluster luminosity and X-ray binary rate \citep{jor07,pea10b}.
The fact that the two GCs that host novae in our survey
are $\sim0.5-1$ mag brighter than the turnover magnitude
is consistent with this mass trend.
We note, however, that because the peak luminosity of novae rival those
of GCs it is very likely that some candidates in the most luminous GCs
may be missed thereby underestimating the dynamical formation rate
and the effect of GC mass
(and the closely connected dynamical interaction rate) on nova formation.

Extragalactic studies of X-ray binaries in GCs have firmly established
that they are three times more abundant in the metal rich population
as compared to the metal poor one in the typical bimodal GC populations
in galaxies, a surprising effect that was hinted at in early Milky Way
and M31 observations \citep{sil75,bel95,kun02}.
The M84 GC nova found here is metal poor while the M87 candidate is metal rich.
Interestingly, based on the published metallicities or colors,
of the 7 other likely or plausible nova GCs only two are likely
to be metal rich. This may hint at the possibility that novae in GCs
are not consistent with the metallicity effect seen in X-ray binaries.
Clearly no strong conclusions can be drawn form this heterogeneous
set of observations with unknown biases.
Future systematic observations such as this study are needed to test
the intriguing possibility that more massive neutron star and black hole
binaries respond differently with metallicity than WD binaries.

\section{Conclusions}

We have reported the results of a Virgo cluster nova survey to test for a
possible relationship between a galaxy's nova rate and its GC
specific frequency. In our CFHT survey spanning a total
of 4 epochs over two years, we have discovered a total of
27, 37, and 19 novae in the fields of M87, M49, and M84, respectively.
These data have led to nova rate estimates of
of $154^{+23}_{-19}$, $189^{+26}_{-22}$,
and $95^{+15}_{-14}$ per year, for the three galaxies.
After considering the
$K$-band luminosities of the galaxies, we find similar
luminosity-specific nova rates, $\nu_K$, for M87, M49, and M84 of
$3.8\pm1.0$, $3.4\pm0.6$, and $3.0\pm0.6$
novae per year per $10^{10}$ $L_{\odot,K}$.
These values are at the high end of the range of $1-3$ novae per year
per $10^{10}~L_{K,\odot}$ found by \citet{gut10}.

The spatial distribution of the M87 novae shows
some evidence that the nova population
could perhaps be separated into two distinct groups:
that of an inner group which follows the background light out
to a distance of about $10'$, similar to what is seen in M49 and M84,
and an outer group of 6 novae clumped between $\sim14'$ and $\sim16'$.
As mentioned previously, 
M87 appears to have an extended halo of diffuse light that is populated
with lower luminosity
elliptical galaxies such as NGC 4478. Thus it is
possible that some or all of the novae seen in the outskirts of M87
may not in fact be associated with M87 itself.
If we restrict our analysis
of M87 to the 21 novae with isophotal
radii less than $10'$, we find
that the cumulative distribution of novae closely follows the background
light distribution.
For this truncated sample, we find
a somewhat smaller nova rate of $124^{+23}_{-17}$ per year for M87,
corresponding to a luminosity-specific nova rate of
$\nu_K=3.1\pm0.8$, which is more in line with the values found for
other galaxies.

We have discovered two novae spatially coincident with known GCs:
one in M87 and one in M84. During the course of our survey a total of
83 novae were discovered in M87, M49 and M84, suggesting that GC
novae could comprise $\sim2.4$\% of the novae seen to erupt in these galaxies.
This estimate is likely a lower limit given the strong observational selection
against the discovery of novae in the brightest clusters (where they may
be most likely to occur). Taken together, the GC systems
comprise only $\sim0.23$\% of the combined masses of the three galaxies.
Thus,
taken at face value, it appears that novae may be enhanced
by at least an order of magnitude in GC environments. Additional monitoring
of the GC systems of Virgo ellipticals, in particular M87,
will be required before any definitive conclusions can be reached.

In summary, the results of our survey
are consistent with the hypothesis that the nova rates in the
three Virgo galaxies studied here are to first order simply
proportional to their mass in stars. Despite the detection of two novae
associated with GCs in M87 and M84 suggesting that
novae may be enhanced in GCs over the field,
there is no compelling evidence that
the overall nova rates in the Virgo ellipticals are particularly sensitive
to the their GC specific frequencies, which vary significantly
between the three galaxies. 

\acknowledgments
We thank an anonymous referee for constructive comments on our original
manuscript.
This work is based on observations obtained with MegaPrime/MegaCam, a joint project of CFHT and CEA/DAPNIA, at the Canada-France-Hawaii Telescope (CFHT) which is operated by the National Research Council (NRC) of Canada, the Institut National des Science de l'Univers of the Centre National de la Recherche Scientifique (CNRS) of France, and the University of Hawaii.
A. W. S. and C. C. acknowledge financial support through NSF grant AST1009566.
C. J. P. acknowledges financial support from the Natural Sciences and
Engineering Research Council of Canada.

\end{document}